\pdfoutput=1
\documentclass[sigconf,nonacm]{acmart}
\usepackage{graphicx} 
\usepackage{xcolor}
\usepackage{xspace}
\usepackage{comment}
\usepackage{verbatim}
\usepackage{tabularx}
\newcommand{\Maike}{\texttt{MAIKE}\xspace}

\title{Thinking Is Not Writing: Designing AI Chatbots to Connect Reasoning and Revision}

\author{Lucile Favero}
\email{lucile@ellisalicante.org}
\affiliation{%
  \institution{ELLIS Alicante}
  \city{Alicante}
  \country{Spain}}

\author{Tanja K\"aser}
\affiliation{%
  \institution{ Swiss Federal Institute of Technology in Lausanne (EPFL)}
  \city{Lausanne}
  \country{Switzerland}}
  
\author{Juan Antonio P\'erez-Ortiz}
\affiliation{%
  \institution{Universitat d'Alacant}
  \city{Alicante}
  \country{Spain}}

\author{Nuria Oliver}
\affiliation{%
  \institution{ELLIS Alicante}
  \city{Alicante}
  \country{Spain}}

\renewcommand{\shortauthors}{Favero et al.}

\date{2026}

\begin{document}

\begin{abstract}
Supporting students' argumentative writing with AI requires connecting reflection to their revision processes. We present \Maike, a Socratic chatbot supporting writing through argument diagnosis, targeted questioning, and a student model across \emph{Think} and \emph{Write} modes. In a classroom study, 65 secondary-school students used \Maike or a control chatbot sharing the interface, language model, and Think/Write modes. 
Approximately 40\% of students improved their essay scores. However, \Maike users engaged more in idea-level dialogue (56\% vs. 44\%) and moved iteratively between discussion and editing. Transitions from ideas discussed in dialogue to revisions were positively associated with score gains. These findings suggest a design target for AI writing support: coordinating reflection with revision. We contribute an architecture for supporting argumentative reasoning without generating essay text and an approach linking interaction patterns to writing processes and outcomes. As AI becomes  embedded in classrooms, understanding these connections is essential to assessing---and designing for---its educational value.
\end{abstract}
\maketitle

\section{Introduction}

Generative AI is rapidly changing how students approach writing \cite{levine2025chatgpt}. Conversational systems can provide explanations, suggestions and help students overcome difficulties during the writing process. However, their ability to generate text also leads to a central challenge for educational writing: support that makes writing easier may reduce the extent to which students themselves engage in the thinking that writing is intended to develop \cite{favero2026substitution}. When an AI system provides an argument, formulates a sentence, or directly writes the essay, students may obtain a better text without having been required to think, articulate, evaluate, or refine the underlying reasoning. 

\begin{figure*}[h!]
    \centering
    \includegraphics[width=\linewidth]{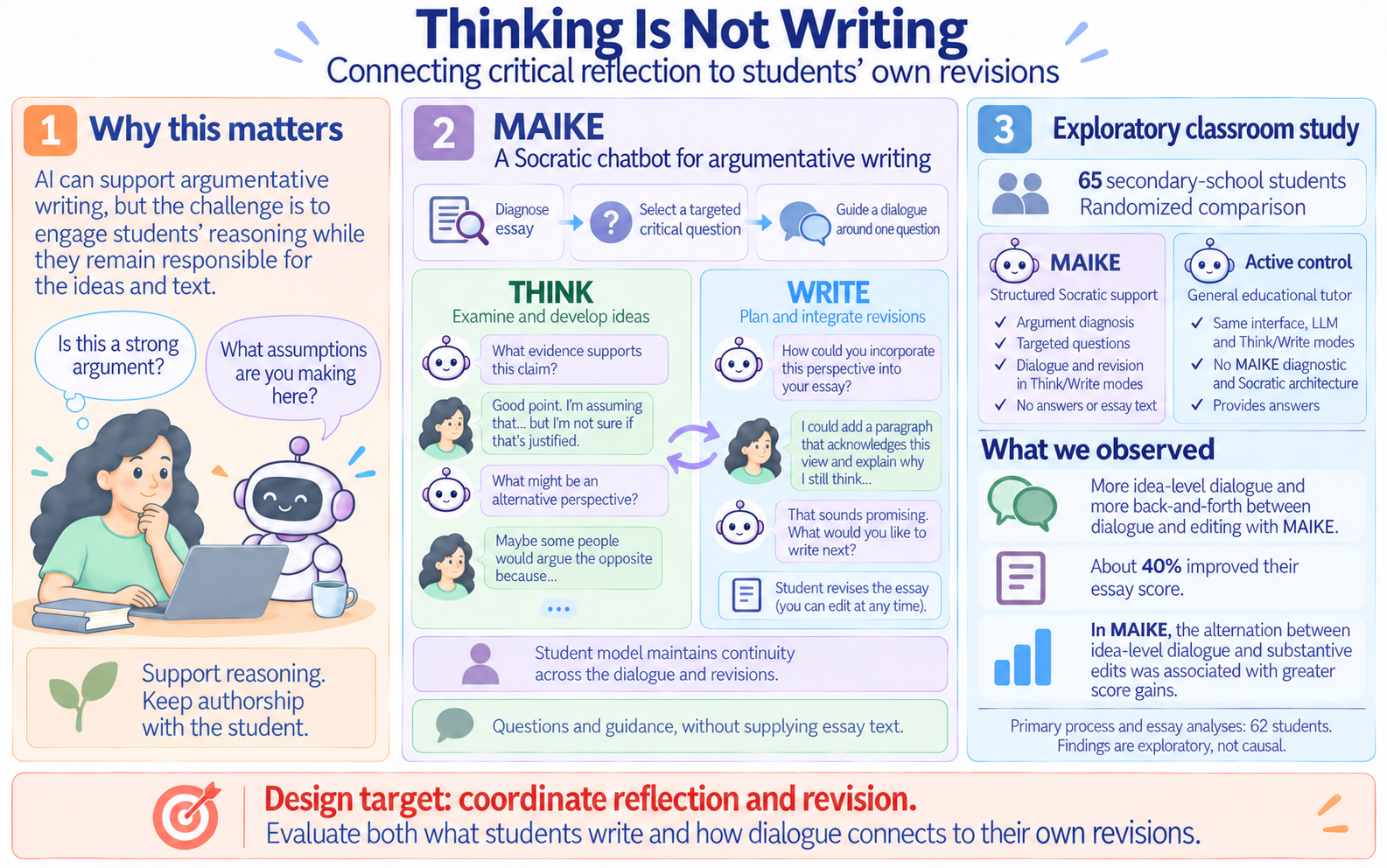}
\caption{Motivation, design, and exploratory classroom study of \Maike. (1) Supporting argumentative writing requires engaging students in reasoning. (2) \Maike identifies weaknesses in an essay and guides students through Socratic dialogue (\emph{Think}) and revision planning (\emph{Write}), without providing answers or writing essay text. (3) A randomized exploratory classroom study with 65 students compares \Maike with a general educational chatbot using the same interface and underlying language model.}
\Description{Three-panel overview showing the motivation for Socratic writing support, the MAIKE cycle from essay diagnosis through questioning to revision, and a randomized classroom comparison with 65 students using the same interface and language model.}
    \label{fig:teaser}
\end{figure*}

This tension is particularly important for argumentative writing. Constructing an argument requires students to formulate and justify claims, examine assumptions, consider evidence and alternative perspectives, and reason about implications \cite{toulmin2003uses,wachsmuth-etal-2017-computational}. These activities are difficult to support through conventional writing assistance because they concern not only what students write but how they arrive at it. Thus, it requires an interaction that helps students develop their own ideas while preserving their responsibility for turning those ideas into text.

Socratic dialogue offers an approach to this problem. Rather than resolving difficulties by providing answers, a Socratic tutor can prompt learners to clarify, justify claims, examine assumptions, consider alternative perspectives, and reason about implications \cite{paul2007critical}. Recent systems have explored this approach through question-based feedback and argument-focused Socratic dialogue for writing \cite{hugenroth2026critical,wang2026question}. However, implementing such an approach in a conversational AI system requires more than prompting a language model to ask Socratic questions. The system must identify aspects of an evolving argument that warrant attention, adapt its questions to the student's responses, maintain continuity across the interaction, and support the transition from reflection to concrete writing and revision.

We present \Maike, an educational chatbot designed around this interaction paradigm. \Maike supports students in developing and revising their own argumentative essays without providing answers or essay text. Its architecture combines argument-focused question selection, a persistent student model, adaptive Socratic response generation, and revision-aware dialogue. \Maike makes an  explicit separation between \textbf{Think} and \textbf{Write} modes. In \emph{Think} mode, \Maike uses targeted Socratic questions to help students articulate and interrogate their reasoning. In \emph{Write} mode, it helps students translate ideas developed through the dialogue into concrete revisions of their own essay. When students resubmit an essay, \Maike reviews the revision in the context of the preceding interaction, allowing subsequent dialogue to build on changes the student made.

We make three contributions. First, we present \textbf{\Maike} as a system for adaptive Socratic writing support, integrating targeted argument-focused questioning, persistent student modeling, and revision-aware dialogue. Second, we introduce \textbf{an interaction-design} approach that separates idea development from textual production, treating the timing and organization of AI assistance as part of the writing activity rather than simply as feedback or text generation. Third, we report an \textbf{exploratory} between-subjects classroom study with 65 secondary-school students comparing \Maike with an active control chatbot. The study is designed to characterize the students' writing processes, explore how they change argumentative writing, and analyze the relationship between interaction trajectories and writing outcomes. Given the modest sample size, we treat the process- and outcome-level comparisons as descriptive groundwork and focus on the process-level implications of the design of educational chatbots that assist students in argumentative writing.
Figure~\ref{fig:teaser} provides a visual summary of the motivation, design, and exploratory classroom evaluation of \Maike.

The remainder of the paper is organized as follows. Section~\ref{sec:related-work} situates our work within previous research on AI-supported writing and Socratic dialogue. Section~\ref{sec:maike} presents \Maike, followed by a description of the classroom exploratory study in Section~\ref{sec:experimental-study}. Section~\ref{sec:findings} examines the students' writing processes, changes in essay quality, and associations between the two. Finally, Section~\ref{sec:discussion} discusses the findings, their implications for the design of AI-based conversational writing support, and the study's limitations.

\section{Related work}

\label{sec:related-work}

\subsection{AI assistance and students' engagement in reasoning}

The educational value of AI writing support depends not only on the quality of the final text, but also on whether students remain engaged in the reasoning that produces it. AI can improve writing while reducing the need for students to formulate, evaluate, and revise their own ideas. In professional writing tasks, users often submitted ChatGPT output with little or no editing \cite{noy2023experimental}. Among students, ChatGPT-supported revisions have likewise been found to improve surface-level errors more than deeper structural problems \cite{pretorius2026mediating}. Greater confidence in generative AI has also been associated with less self-reported critical thinking \cite{lee2025impact}. AI assistance can also influence the positions writers develop: \citet{jakesch2023co} found that co-writing with an opinionated language model shifted both the views expressed in participants' texts and their subsequently reported opinions. 
\citet{tankelevitch2024metacognitive} further argue that using generative AI places demands on users' ability to monitor and regulate their thinking, and that interfaces should support these processes. This perspective motivates guidance that helps students examine their understanding and decide how to act on feedback.

These findings raise a central design question: \emph{how can AI support the writers without taking over their reasoning?}
This question is the theoretical starting point of \Maike,
supported by a taxonomy of 
four interconnected dimensions impacted by AI: \emph{cognition}, through cognitive offloading and superficial learning; \emph{agency}, through dependency, overtrust, and intellectual conformity; \emph{emotional well-being}, including lower self-efficacy and stress; and \emph{ethics}, including privacy and academic integrity
 ~\cite{wood1976scaffold,favero2026substitution}.
\Maike mainly addresses the \emph{cognitive} and \emph{agency} dimensions from that taxonomy. It does not provide answers or essay text, but asks students to explain, examine, and revise their own reasoning. 
The student remains responsible for deciding whether it matters, what they think, and what enters the essay. 
We refer to this approach as preserving the student's \emph{epistemic responsibility}.

The ICAP (Interactive–Constructive–Active–Passive) framework provides a complementary perspective by distinguishing forms of engagement based on what learners do \cite{chi2014icap}. A long conversation with an AI system may involve little reasoning if the student mainly accepts suggestions, while a shorter exchange may be productive if it leads them to justify or reconsider an idea. \citet{levine2025chatgpt}, for example, found that high-school students used ChatGPT differently across writing stages. We therefore examine not only how much students interact with the chatbot, but also what kind of reasoning they contribute and how this connects to revision.

\subsection{Argument-focused writing support}

Argumentative writing requires attention to claims, evidence, reasoning, and alternative perspectives. Argument mining provides methods for identifying argumentative components and their relations \cite{stab-gurevych-2017-parsing}, including a distinction between the logical, rhetorical, and dialectical dimensions of argumentative quality \citet{wachsmuth-etal-2017-computational}. These methods can help identify weaknesses, but the pedagogical question is how to use this analysis without replacing the student's reasoning.
ArgueTutor combines argument analysis with conversational adaptive feedback \cite{wambsganss2021arguetutor}, while VISAR combines visual representations of argument structure with AI support for exploring and developing arguments \cite{zhang2023visar}. Both show how argument analysis can support writing.

\Maike uses argument analysis differently. It analyzes the essay internally to identify issues worth discussing, but does not show the student a score, correction, or ready-made solution. Instead, the analysis determines what the system asks about. The student must then examine the issue and decide whether and how to revise the essay.

\subsection{Socratic dialogue for argumentative writing}

Socratic dialogue by means of critical questions provides a natural interactional form to support argumentative writing. For example, Paul and Elder's framework of critical questioning describes questions that clarify ideas, probe assumptions and evidence, examine alternative perspectives, and explore implications \cite{paul2007critical}. Rather than telling students what to think or write, such questions make their reasoning explicit and open it to examination.

In HCI, \citet{danry2023don} found that framing AI feedback as questions helped participants identify logically flawed statements more accurately than providing direct explanations. This supports questioning as an interaction approach for engaging users' reasoning, although its role in sustained writing and revision requires further study.

Recent systems have proposed Socratic questioning to writing support. \citet{wang2026question} studied a question-only Socratic assistant for second-language argumentative writing. Compared with conventional peer and teacher feedback, students showed greater gains in evidence use and counterargument integration, but not in stance. Its questions, however, came from a predefined bank rather than being generated dynamically and in real-time from the student's essay. \citet{namaziandost2026socratic} similarly reported larger gains in critical thinking, epistemic beliefs, and argumentation with a Socratic agent than with a generic chatbot or no AI.
Critical Inker is closer to \Maike \cite{hugenroth2026critical}. It combines argument extraction and evaluation with a Socratic chatbot that asks targeted questions and marks planned revisions in the text after students explain what they want to change. \Maike similarly connects argument-grounded dialogue with revision, but makes this transition part of the conversation: in \emph{Think} mode, students examine the issue whereas in \emph{Write} mode, they translate that reasoning into a concrete revision.

\Maike builds on prior work on these areas, including argument mining and automated assessment of argumentative writing, Socratic dialogue generation, and critical-question generation and selection, combining these approaches  to guide students in their argumentative writing: it identifies which issue to address, adapts the dialogue to the student's reasoning across turns, and helps the student translate reflection into revision.

\subsection{Dialogue, revision, and writing outcomes in argumentative essay writing}

Evaluating whether AI supports students' reasoning requires examining how they use it, not only the quality of their final essays. In fact, receiving feedback does not necessarily lead to revision: \citet{jansen2025feedback} found that in logs from more than fourteen thousand students, around half made no revision after receiving AI feedback. Students may also reject content-focused feedback when they believe the system has misunderstood their meaning \cite{chen2026barriers}. These findings highlight the importance of understanding how students interpret feedback and decide whether to act on it. For conversational writing support, this means examining both the reasoning students express in dialogue and how they carry it into their revisions.
Writing is also not a fixed sequence of stages. \citet{flower1981cognitive} describe \emph{planning}, \emph{translating}, and \emph{reviewing} as processes that can recur throughout writing, while \citet{faigley1981analyzing} distinguish surface revisions from changes that affect meaning. This motivates examining how students move between discussing ideas and making substantive revisions.
Keystroke logs have been utilized to characterize writing and revision strategies \cite{zhu2019keystroke,mouchel2023revision}. Process mining has more broadly been used to identify learning strategies from traces of learner activity \cite{matcha2019analytics}. 
In human--AI writing, the CoAuthor system similarly recorded keystroke-level interaction with GPT-3 suggestions \cite{lee2022coauthor}. Socratic writing studies, however, have focused more on outcomes than on the process. For example, \citet{wang2026question} recorded dialogue turns but did not conduct fine-grained process tracing. Less is known about how idea-level dialogue connects to substantive essay revision.
This matters because different forms of feedback do not always lead to different final texts. \citet{meyer2026comparing} found no significant differences in revision performance between teacher-written and LLM feedback conditions, while \citet{farrokhnia2026generative} found that higher-rated AI feedback did not produce better revised essays after controlling for the initial draft. A tutoring design may therefore change \emph{how} students reason and revise even when differences in final essay scores are limited.

Our exploratory classroom study evaluates \Maike as an implementation of the \emph{scaffold, do not substitute} principle \cite{favero2026substitution}: 
it provides argument diagnosis, targeted question generation, student modeling, response selection, and revision review. We examine whether this structure changes the students' reasoning and revision processes, whether it leads to different changes in argumentative writing, and whether moving between idea-level discussion and substantive revision is associated with final essay improvement.

\section{\Maike: a Socratic educational chatbot}
\label{sec:maike}
This section first presents the design goals of \Maike (Section~\ref{sec:maike-goals}) and the student's interaction with the system (Section~\ref{sec:maike-overview}). Next, it describes how the essay is analyzed (Section~\ref{sec:diagnosis}), how this analysis informs critical-question selection (Section~\ref{sec:maike-questions}), and how the tutor adapts the dialogue to the student's contributions (Section~\ref{sec:maike-dialogue}). Finally, it explains how revisions inform subsequent interaction (Section~\ref{sec:maike-revisions}) and provides an example (Section~\ref{sec:maike-example}) and implementation and reproducibility details (Section~\ref{sec:maike-implementation}).

\subsection{Design goals and principles}
\label{sec:maike-goals}
\Maike is a web-based tutoring system designed to support secondary-school students as they develop and revise short argumentative essays. Its central design goal is to help students examine their reasoning while retaining responsibility for the ideas and wording of their text. To this end, the tutor is instructed to guide students through questions, without providing substantive answers, sample arguments, or rewritten passages.

Three principles organize this support. First, \textbf{ground the dialogue in the student's argument}: questions should address issues identified in the submitted essay, including weaknesses in specific passages and in the argument as a whole. Second, \textbf{respond to the student's evolving contribution}: subsequent questions should build on what the student has said, clarify difficulties, and maintain continuity across the conversation. Third, \textbf{connect reflection to revision}: the interaction should help students consider how ideas developed through dialogue could change their essay, while leaving them to decide and write those changes.

These principles define the intended interaction. \Maike does not retrieve external information, verify the factual accuracy of claims, or provide grammar correction. Its internal analyses guide the conversation and are not presented to students as scores or corrective feedback. The tutor may rephrase or explain its own questions when needed, but is instructed not to answer the substantive issue under discussion.

\subsection{Student interaction and system overview}
\label{sec:maike-overview}

The interface combines an essay editor with a conversational panel. Figure~\ref{fig:sys-overview} illustrates the repeated analysis--dialogue--revision cycle:
\begin{figure*}[h!] 
    \centering
    \includegraphics[width=0.9\linewidth]{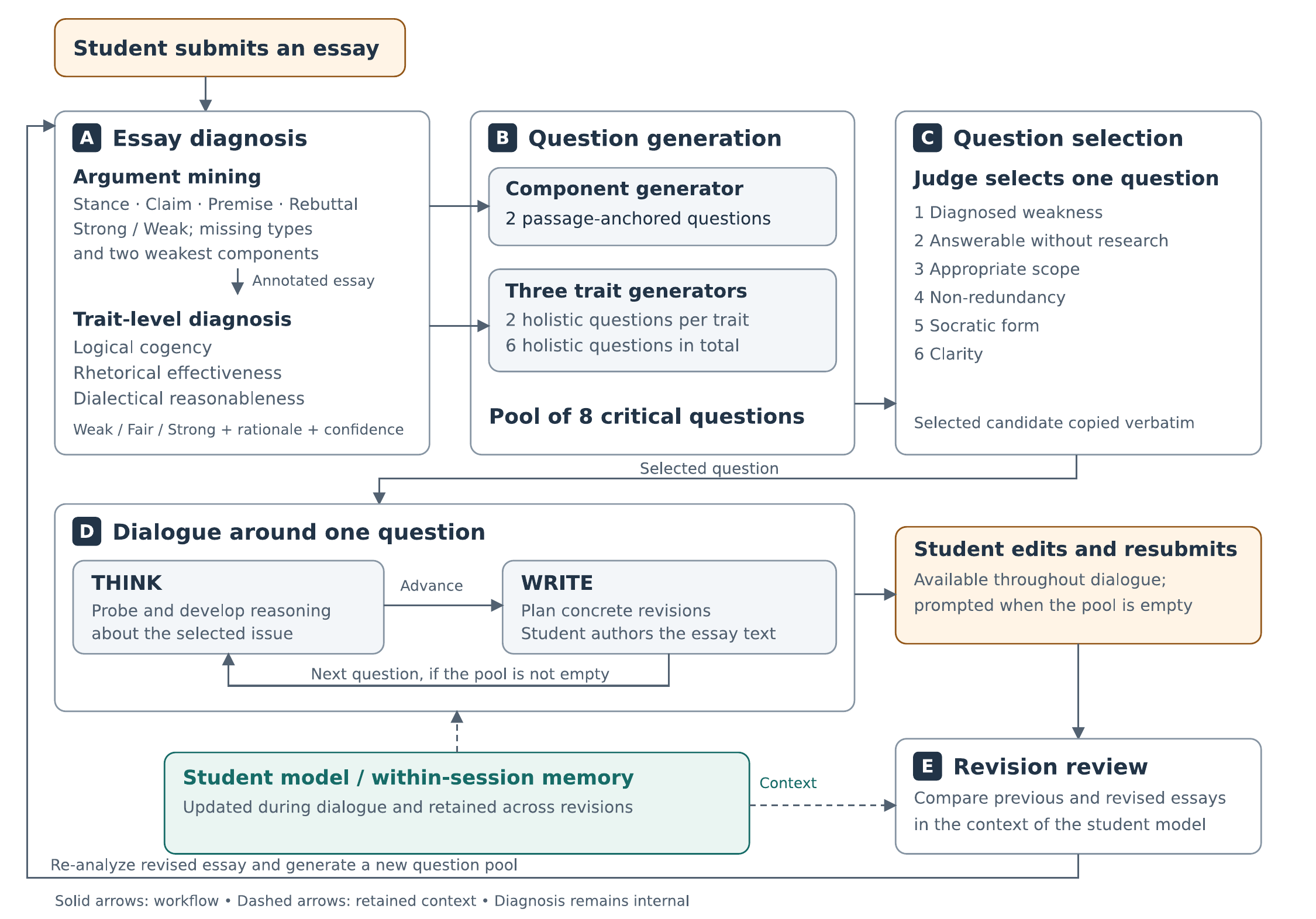}
    \caption{\Maike’s analysis–dialogue–revision cycle. (A) Component- and trait-level diagnosis informs (B) four generators producing eight critical questions. (C) A Judge selects one question. (D) \emph{Think} mode probes the selected issue, while \emph{Write} mode supports revision planning. Students can edit and resubmit throughout the dialogue. (E) Revision review compares successive essays using the retained student model before restarting diagnosis and question generation. Dashed arrows indicate contextual support from within-session memory.}
\Description{Flowchart linking essay diagnosis, generation of eight critical questions, selection of one question, Think-mode discussion, Write-mode revision planning, and revision review. Resubmission restarts the cycle. Dashed arrows show contextual support from a student model retained across dialogue turns and revisions.}
    \label{fig:sys-overview}
\end{figure*}

\begin{enumerate}
    \item The student submits a short argumentative essay.

    \item \Maike analyzes the essay's argument structure and its overall argumentative quality (\textbf{Module A}).

    \item From this diagnosis, four specialized generators produce a pool of eight critical questions targeting the identified weaknesses (\textbf{Module B}). A separate Judge selects one question from this pool (\textbf{Module C}).

    \item The conversation begins with the selected question. In \emph{Think} mode, \Maike asks follow-up questions to help the student clarify, justify, and examine their reasoning about the identified issue (\textbf{Module D}).

    \item In \emph{Write} mode, \Maike helps the student plan how to translate the discussion into a concrete revision of the essay (\textbf{Module D}). The student decides what to change and writes the revised text; the tutor does not supply essay wording.

    \item \Maike then moves to another question from the pool and repeats the Think--Write sequence (\textbf{Module D}). When the pool is empty, the student is prompted to revise and resubmit the essay.

    \item At any point during the conversation, the student can edit and resubmit their essay, regardless of the current mode.

    \item Upon resubmission, \Maike compares the previous and revised essays in the context of the preceding interaction (\textbf{Module E}). It then analyzes the revised essay and generates a new question pool, starting another cycle (\textbf{Modules A--C}).
\end{enumerate}

The student model is updated throughout the dialogue and retained across revision cycles. As shown by the dashed links in Figure~\ref{fig:sys-overview}, it provides context for both the dialogue and the revision review, allowing subsequent interaction to build on what the student has discussed and changed. The following subsection describes the role and output of each component.


\subsection{Essay diagnosis (Module~A)}\label{sec:diagnosis}
\paragraph{Argument mining}
To protect the students' privacy, the argument-mining step in Module~A of Figure~\ref{fig:sys-overview} is informed by prior work on argument mining with small, locally run language models \cite{favero2026leveraging}. A prompted model reproduces the essay with argumentative spans tagged as \textit{Stance}, \textit{Claim}, \textit{Premise}, or \textit{Rebuttal}. In the system's label set, a stance represents the essay's main position, a claim supports that position, and a premise supports a claim or an opposing argument. The label \textit{Rebuttal} introduces reasoning from an opposite or alternative viewpoint \cite{toulmin2003uses,stab-gurevych-2017-parsing}.

Each identified component receives a binary quality label (\textit{Strong} or \textit{Weak}). The same LLM call produces (1) the essay tagged with component type and quality, (2) a short overall justification, (3) any component types identified as missing, and (4) the two components judged the weakest. Note that the resulting annotated representation is an internal diagnostic used by subsequent modules. It is not displayed to the student.

\paragraph{Essay-level traits}

Three separate prompted calls evaluate the annotated essay along argumentative quality dimensions proposed by \citet{wachsmuth-etal-2017-computational}: \textit{logical cogency}, \textit{rhetorical effectiveness}, and \textit{dialectical reasonableness}. \textit{Logical cogency} concerns the acceptability, relevance, and sufficiency of the reasoning. \textit{Rhetorical effectiveness } concerns the clarity, organization, tone, credibility, and persuasive presentation of the argument. \textit{Dialectical reasonableness} concerns whether the essay engages fairly and coherently with alternative positions.
For each trait, the model produces (1) a diagnostic label  \textit{Weak}, \textit{Fair}, or \textit{Strong}, (2) a justification, and (3) a confidence level.   Because the writing task excludes external sources, the prompts instruct the model not to penalize the absence of citations. These outputs are used internally to guide question generation; the study's essay outcomes are assessed separately by human raters.

\paragraph{Complementarity of the two analyses}
The component- and trait-level analyses provide two complementary diagnoses of the essay. Component-level diagnoses identify local weaknesses associated with specific passages, whereas trait-level diagnoses identify deficiencies in the text as a whole. This distinction directly structures the next stage: component diagnoses generate locally anchored questions, while trait diagnoses generate holistic questions.

\subsection{From diagnosis to a critical question (Module~B-C)}
\label{sec:maike-questions}

\Maike uses the validated question generation architecture developed by \citet{favero-etal-2025-ellis}. 
First, the \emph{Questioner} produces a diverse pool of critical questions, using several specialized generators. Each one draws on a different source of information, theory, or analysis. Second, the \emph{Judge} selects the most appropriate question according to predefined criteria. In the \Maike system, the generators draw on the different components of the essay diagnosis (see Module~B in Figure~\ref{fig:sys-overview}). 
More specifically, one generator targets weak argument components identified by the argument-mining analysis. Their questions require quoting the relevant part of the essay. Three additional generators target the essay-level traits (logical cogency, rhetorical effectiveness, and dialectical reasonableness), with one generator dedicated to each trait.
These generators produce holistic questions without quoting specific passages.
Module~C in Figure~\ref{fig:sys-overview} implements the \emph{Judge} as a separate prompted call that selects one candidate from the pool. It prioritizes, in order, relevance to a diagnosed weakness, answerability without external research, appropriate scope, non-redundancy, Socratic form, and clarity. The selected question provides the focus for the ensuing dialogue.

\subsection{Adaptive Socratic dialogue (Module~D)}
\label{sec:maike-dialogue}

\begin{figure*}
    \centering
    \includegraphics[width=\linewidth]{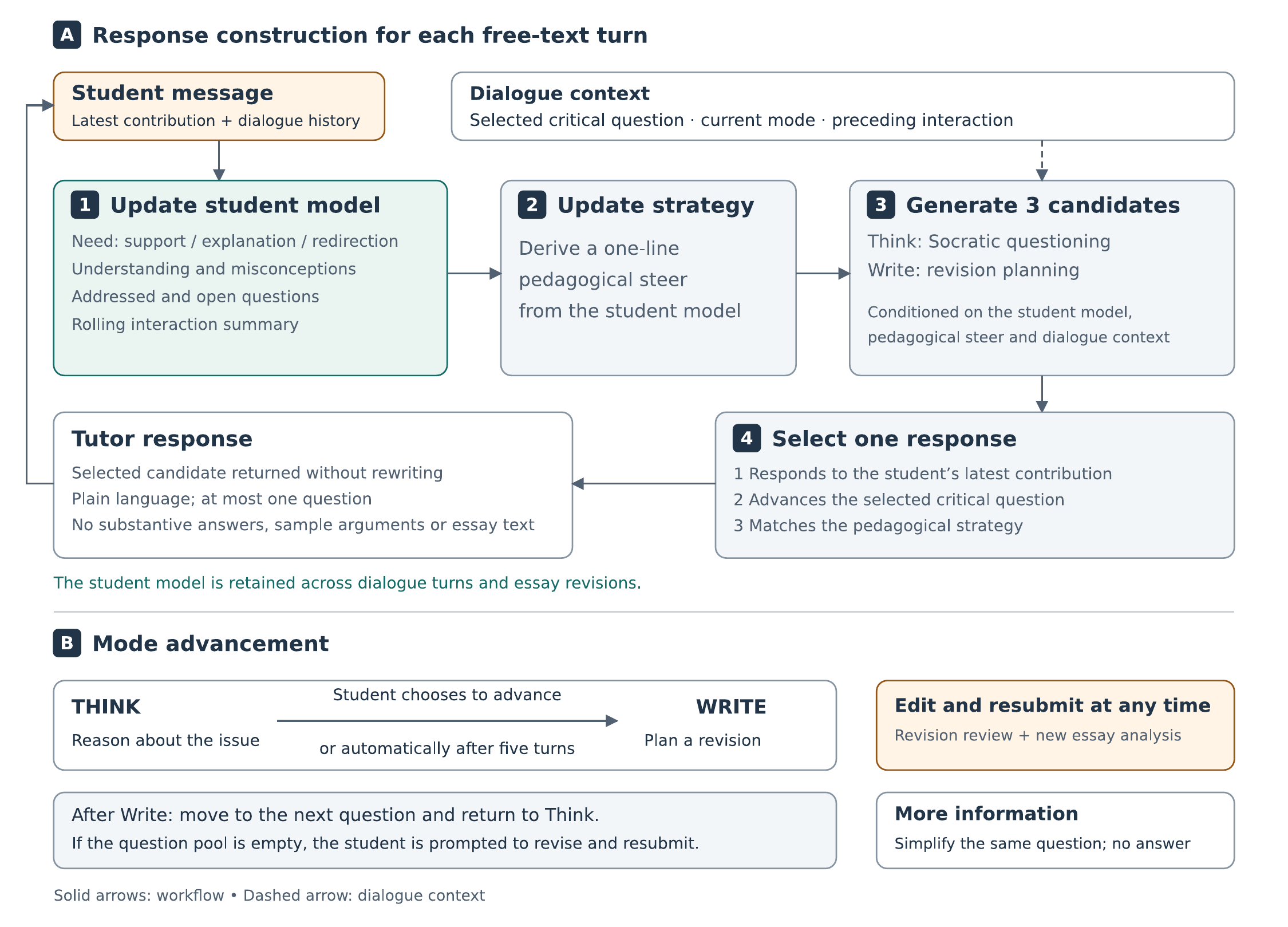}
    \caption{\Maike’s dialogue logic. (A) For each free-text student turn, Module~D (1) updates the student model, (2) derives a pedagogical strategy, (3) generates three candidate responses, and (4) selects one. The selected response must respond to the student's latest contribution, advance the selected critical question, and follow the pedagogical strategy. (B) Each critical question is explored first in \emph{Think} mode and then in \emph{Write} mode. Students can advance explicitly, or the system advances after five conversational turns in the current mode, excluding elaboration requests. After Write, the dialogue moves to another question when available. Students can edit and resubmit their essay at any time.}
\Description{Two-panel flowchart. Panel A shows student-model updating, pedagogical strategy selection, generation of three candidate responses, and selection of one response. Panel B shows progression from Think to Write, then to another question. Students can request clarification or edit and resubmit at any time.}
    \label{fig:dia_logic}
\end{figure*}

Once Module~C has selected a critical question, Module~D turns this question into an adaptive dialogue that connects reasoning to revision. Figure~\ref{fig:dia_logic} summarizes the two parts of this process: Panel~A shows how \Maike constructs a response to each typed student turn, and Panel~B shows how the dialogue progresses through \emph{Think} and \emph{Write} modes.

\subsubsection{Conversational progression}

Panel~B of Figure~\ref{fig:dia_logic} shows how the dialogue progresses through two student-visible modes for each critical question selected by Module~C. In \emph{Think} mode, \Maike helps the student examine the issue raised by the question by clarifying, justifying, or further developing their reasoning. In \emph{Write} mode, it helps the student translate the ideas developed in the dialogue into a concrete plan for revising the essay. In each mode, students can choose to advance using the interface. Otherwise, \Maike advances after five conversational turns. This limit applies separately to \emph{Think} and \emph{Write}, and elaboration requests do not count toward it. After \emph{Write}, \Maike moves to another question from the pool and returns to \emph{Think} mode. When no questions remain, the student is prompted to revise and resubmit the essay. A resubmission activates Module~E and subsequently starts a new analysis and question-generation cycle through Modules~A--C. Students can also edit and resubmit their essay at any point during the dialogue.

This structure supports two connected needs: examining an idea and deciding how to integrate it into the essay. The distinction between \emph{Think} and \emph{Write} draws on work on the interaction between content reasoning and rhetorical planning \cite{scardamalia1984teachability}. 
Working one focused question at a time through a Socratic conversation that splits into \emph{Think} and \emph{Write} modes makes the task more manageable, following scaffolding principles \cite{wood1976scaffold}. Allowing students to edit and resubmit throughout the dialogue also reflects the recursive nature of writing \cite{flower1981cognitive}: examining ideas can lead to revisions, while revising can reveal new issues to explore.

The interface also provides an \emph{elaboration} option, which rephrases or simplifies \Maike's latest message while preserving the same underlying question. It is constrained not to provide the answer or introduce new substantive content. This option applies Paul and Elder's emphasis on clarity to the tutor's own questions: students can request a clearer formulation before responding \cite{paul2007critical}.

 \subsubsection{Response construction}

Panel~A of Figure~\ref{fig:dia_logic} shows how Module~D constructs a response to each typed student message. The process consists of four steps, numbered consistently in the figure and below.

\paragraph{1. Student-model update}

In Step~1 of Figure~\ref{fig:dia_logic}, \Maike interprets the student's latest message in the context of the preceding interaction and updates its student model. This model captures four elements: (1) the student's immediate need, classified as \emph{support}, \emph{explanation}, or \emph{redirection}; (2) their readiness to progress in the current mode; (3) their apparent understanding and possible misconceptions; and (4) a summary of the interaction so far (including which issues have been addressed, which remain open, and a rolling summary of the preceding dialogue). These quantities are model-generated inferences used only to guide the subsequent dialogue.

The three students' immediate need categories motivate different forms of assistance: \emph{support} helps students work through difficulty, \emph{explanation} clarifies what is being asked, and \emph{redirection} restores focus. This distinction is consistent with  D'Mello and Graesser's research on affective state changes during learning, which motivates attending to confusion and disengagement as dialogue unfolds \cite{d2012dynamics}. 

\paragraph{2. Pedagogical-strategy update}

In Step~2, \Maike uses the updated student model to derive a one-line pedagogical steer specifying the intended function of the next response. The steer  connects the inferred state of the student to the response-generation step rather than directly determining the wording of the tutor's reply. This follows the scaffolding principle of adjusting guidance to the learner's needs while leaving them responsible for the reasoning \cite{wood1976scaffold}.

\paragraph{3. Candidate-response generation}
In Step~3, \Maike generates three alternative responses conditioned on the student model, pedagogical steer, selected critical question, current mode, and preceding dialogue (Figure~\ref{fig:dia_logic}, A). Generating several candidates separates response generation from selection.

\begin{figure*}[t]
    \centering
    \includegraphics[width=0.8\textwidth]{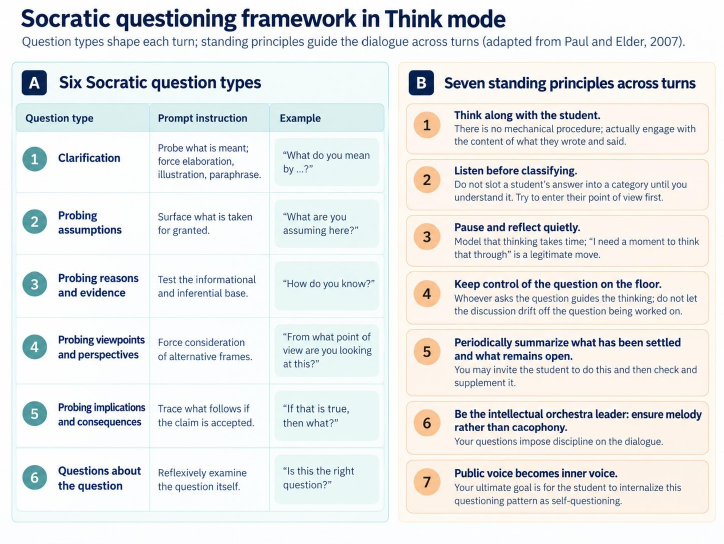}
    \caption{Socratic questioning framework used in \Maike's Think mode. 
    (A) Six question categories determine the type of Socratic question generated on each turn. 
    (B) Seven standing principles guide how the dialogue develops across turns.}
    \Description{Two-column framework presenting six question types: clarification, assumptions, reasons and evidence, viewpoints, implications, and questions about the question. Seven dialogue principles emphasize engagement, listening, reflection, focus, summarizing, coherence, and student self-questioning.}
    \label{fig:socratic-framework}
\end{figure*}

In \emph{Think} mode, each candidate question is guided by Paul and Elder's Socratic principles \cite{paul2007critical} and is built as one of six Socratic-question categories, selected before the question is generated and constraining the form of the response. Beyond the individual question, seven standing Socratic principles guide how the dialogue develops across turns, instructing \Maike to engage with the student's actual reasoning before responding to it, allow space for reflection, maintain focus on the question currently being examined, and periodically consolidate what has been established and what remains unresolved, with the longer-term aim of helping students internalize this questioning process as self-questioning. Figure~\ref{fig:socratic-framework} summarizes both components of this \emph{Think}-mode questioning framework; the full category and principle definitions, with prompt examples, are reproduced in the supplementary material.

In \emph{Write} mode, the candidates instead focus on turning the reasoning developed in \emph{Think} mode into a concrete revision plan. They may identify what the student could clarify, qualify, connect, or develop, but they do not provide the wording of the revision.

Across both modes, every candidate-response is constrained to (1) explicitly connect to the student's latest contribution, (2) remain focused on the selected issue, (3) use plain language, (4) contain  one question, and (5) avoid substantive answers, sample arguments, or proposed essay sentences. Note that these constraints operationalize the distinction between supporting the student's reasoning and supplying the content of the essay.

\paragraph{4. Response selection}

Finally, Step~4 of Figure~\ref{fig:dia_logic} uses a separate prompted call to select one of the three candidates. The selector receives the candidates together with the student's latest contribution, the selected critical question, and the current pedagogical steer. It prioritizes candidates that (1) visibly respond to what the student has just said, (2) advance the selected critical question, and (3) follow the pedagogical strategy.

\subsection{Revision review and within-session continuity (Module~E)}
\label{sec:maike-revisions}

Module~E in Figure~\ref{fig:sys-overview} handles revision review. When a student resubmits the essay, a revision-review call compares the previous and revised versions in the context of the student model. Its output allows subsequent dialogue to refer to changes the student has made and to issues discussed earlier. The return arrow from Module~E to Module~A represents the start of a new cycle: the revised essay passes through diagnosis, question generation, and selection again (Modules~A--C).

The student model is retained across these revision cycles within the session, as shown by the memory block and dashed contextual links to Modules~D and~E in Figure~\ref{fig:sys-overview}. It provides conversational continuity. Its contents are model-generated inferences, not validated assessments of the student's knowledge or affect. 

\subsection{Example of an interaction with \Maike}
\label{sec:maike-example}
\begin{figure*}[h!]
\centering
\includegraphics[width=\textwidth]{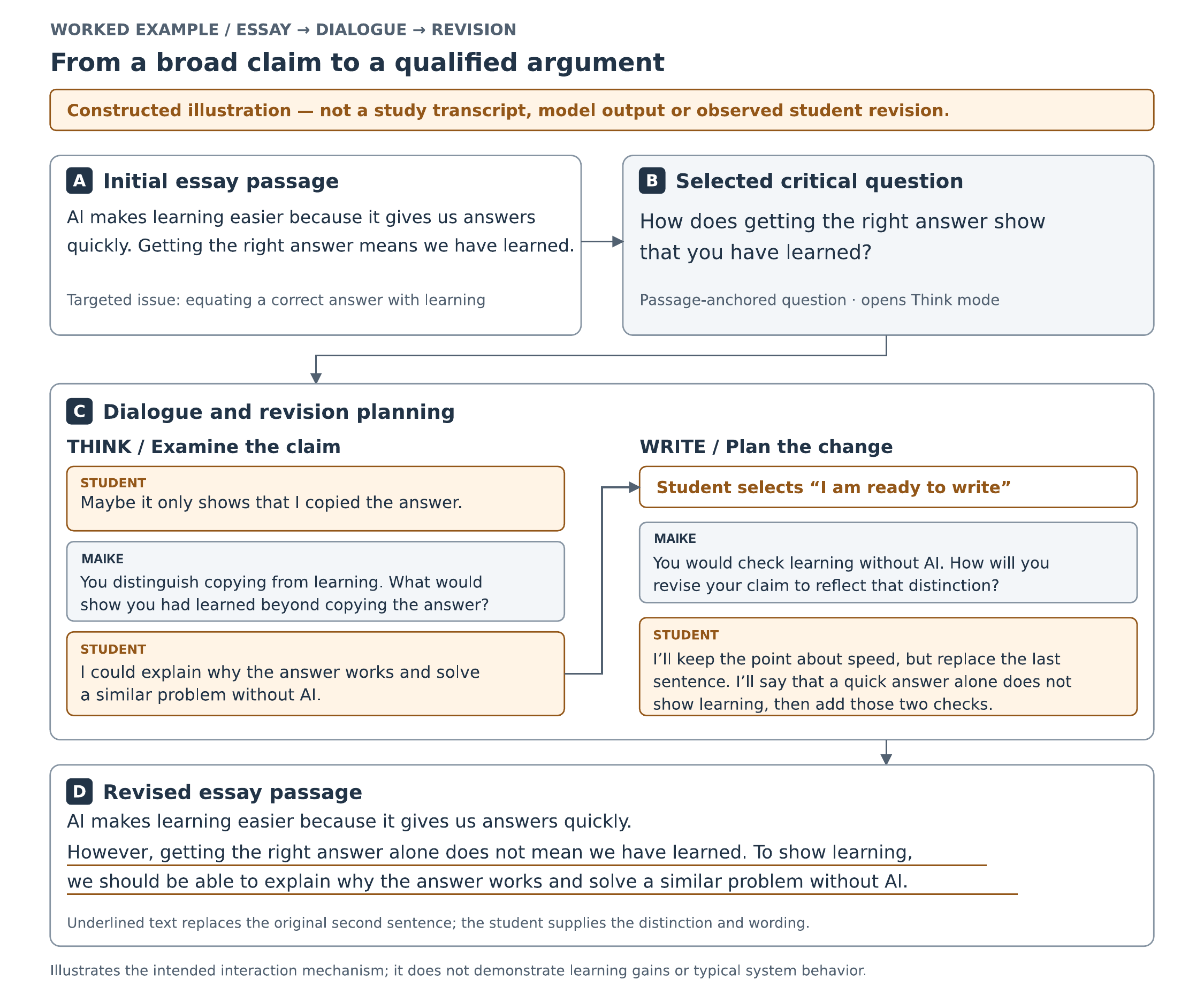}
\caption{Example of \Maike's intended interaction. (A) The initial passage about obtaining a correct answer with learning. (B) \Maike asks a question targeting this assumption. (C) The Socratic dialogue and revision Planning. (D) The revised passage qualifies the original claim. Underlining marks the revised text. The example is illustrative and was not taken from the study.}
\Description{Four-part illustrative example showing how dialogue can support essay revision. Panel A presents a passage equating getting the correct answer with learning. Panel B questions this assumption. Panel C shows the student distinguishing copying from understanding in Think mode, then planning how to revise the claim in Write mode. Panel D presents the revised passage. This constructed example is not taken from the study.}
    \label{fig:worked-example}
\end{figure*}

Figure~\ref{fig:worked-example} illustrates how \Maike connects dialogue about the student's argument to a concrete revision of the essay.

(A) The example starts with a passage from a student's essay about ``obtaining the correct answer as evidence of learning'' . (B) \Maike asks a targeted question that prompts the student to examine this assumption. In this example, the selected question was generated from one of the weakest argument components identified by the argument-mining analysis and was therefore anchored to this specific passage. It was then selected from the pool of generated critical questions. (C) In \emph{Think} mode, the student develops a distinction between copying an answer and understanding it. In \emph{Write} mode, the student then considers how to integrate this distinction into the essay. Throughout the dialogue, the student provides the ideas and the revised text, while \Maike guides the process through questions. (D) Finally, the student revises the essay to incorporate the distinction developed during the dialogue. This is a constructed example of the intended interaction, not an exchange observed in the study.

\subsection{Implementation and reproducibility}
\label{sec:maike-implementation}
\Maike consists of an Angular\footnote{\url{https://angular.dev/}} 19 / Ionic\footnote{\url{https://ionicframework.com/}} 8 client and a FastAPI\footnote{\url{https://fastapi.tiangolo.com/}} backend, using Ministral 3 8B Reasoning 2512 \cite{liu2026ministral}---a small, open-weight model chosen to limit computational cost and enable institution-controlled hosting. The model ran on a research cluster. Each pipeline task is a separate, non-fine-tuned model call using a fixed prompt template and in-context learning \cite{brown2020language, dong2024icl, wu2022aichains, khot2023decomposed}, with JSON-schema validation and regeneration on invalid output, and rationale-before-label chain-of-thought reasoning \cite{wei2022cot, kojima2022zeroshot, tam2024format}.  The complete prompts for both study conditions are in the supplementary material.

Task behavior was specified entirely in the prompt through in-context learning~\cite{brown2020language, dong2024icl}. Most tasks are zero-shot, relying on the definition section for the label inventory and the criteria for applying it. Only argument mining is one-shot, carrying a single worked essay-to-annotation pair that fixes the output format and the label space \cite{min2022rethinking}. Outputs were restricted to a single JSON object with a fixed key set and closed label vocabularies. This restriction was not enforced through constrained decoding \cite{geng2023gcd}: responses were checked against task-specific schemas after generation, and an invalid response triggered a regeneration, with up to six attempts per task. Within that object, every task producing a judgment returned its rationale before the label it supports, a structured form of chain-of-thought prompting \cite{wei2022cot, kojima2022zeroshot} that also mitigates the loss in reasoning quality that output-format restrictions can introduce \cite{tam2024format}.
The supplementary material includes the complete prompts for both study conditions.

\section{Exploratory classroom study}
\label{sec:experimental-study}

To examine what \Maike's structured Socratic interaction contributes beyond general educational chatbot support, we conducted an exploratory randomized between-subjects classroom study in which students used either \Maike or a control chatbot to write an argumentative essay. We first characterize \emph{how} students interacted with the chatbots and \emph{how their argumentative writing changed} from initial to final essay. We then explore whether differences in how students coordinated idea-level reasoning with essay revision are associated with changes in writing. We address three research questions: 

\begin{itemize}
    \item[RQ1.] How do students' interactions differ between \Maike's diagnostic Socratic approach and general educational chatbot support, particularly in how students move between idea-level discussion and essay revision?
    \item[RQ2.] How does the students' argumentative writing change from their initial to final essays, and how do these changes compare across conditions? 
    \item[RQ3.] How is the coordination of idea-level discussion and essay revision associated with changes in the students' argumentative writing?
\end{itemize} 

\subsection{Participants, recruitment and ethics}
Participants were recruited from four Theory of Knowledge (ToK) class cohorts, each at a different
Swiss secondary school offering the International Baccalaureate program.   In the ToK class, students are required to develop and write argumentative essays. Teachers distributed information sheets to students and their parents. These were also available through the in-app consent page and described the study and its potential benefits for students’ learning. Participation was voluntary and uncompensated. Students and their parents could request the withdrawal of the student’s data from research use for up to one month after the study session. The study received approval from the relevant Human Research Ethics Committee of EPFL.

A total of $67$ students took part in the study and generated data. Two students did not consent to the use of their data for research, resulting in a final analysis sample of $65$ students: $34$ in the control condition and $31$ in the \Maike condition. Demographic information was available for $63$ participants as two did not complete the demographics questionnaire. Of those who reported their gender, $32$ identified as female, $26$ as male and one as non-binary. Four participants preferred not to report their gender. Participants were between $12$ and $19$ years old ($M=17.0$, median $=17$). Additional demographic information is provided in the supplementary material.  

The study was conducted in Switzerland between June and July 2026. Participant records were pseudonymized and stored on a controlled service compliant with existing data protection legislation.
The study was conducted entirely in English, which only five of the 63 students named as a first language. For the others, it was a second language. However, they had been learning English for $9.2 \pm 2.0$ years on average, with a mean self-reported English grade of $5.2 \pm 0.6$ on the Swiss 1--6 scale, and they took the ToK class in English. 

The \Maike and control conditions were similar on all six pre-test attitude scales (Cliff's $\delta$ ranged from $-.17$ to $+.03$, with all 95\% intervals including zero). Students in the \Maike condition reported slightly lower overall grades on average than those in the control condition ($4.99$ vs.\ $5.09$ on the Swiss 1--6 scale) (see supplementary material). 

The sample size was determined by school and classroom availability rather than
an a priori power analysis. The modest sample limits the precision of
our estimates. A non-significant result should therefore not be
interpreted as evidence that the conditions are equivalent and the study should be interpreted as exploratory.

\subsection{Experimental design, conditions and assignment}
\label{sec:experimental-design}
\begin{figure*}
    \centering
    \includegraphics[width=\linewidth]{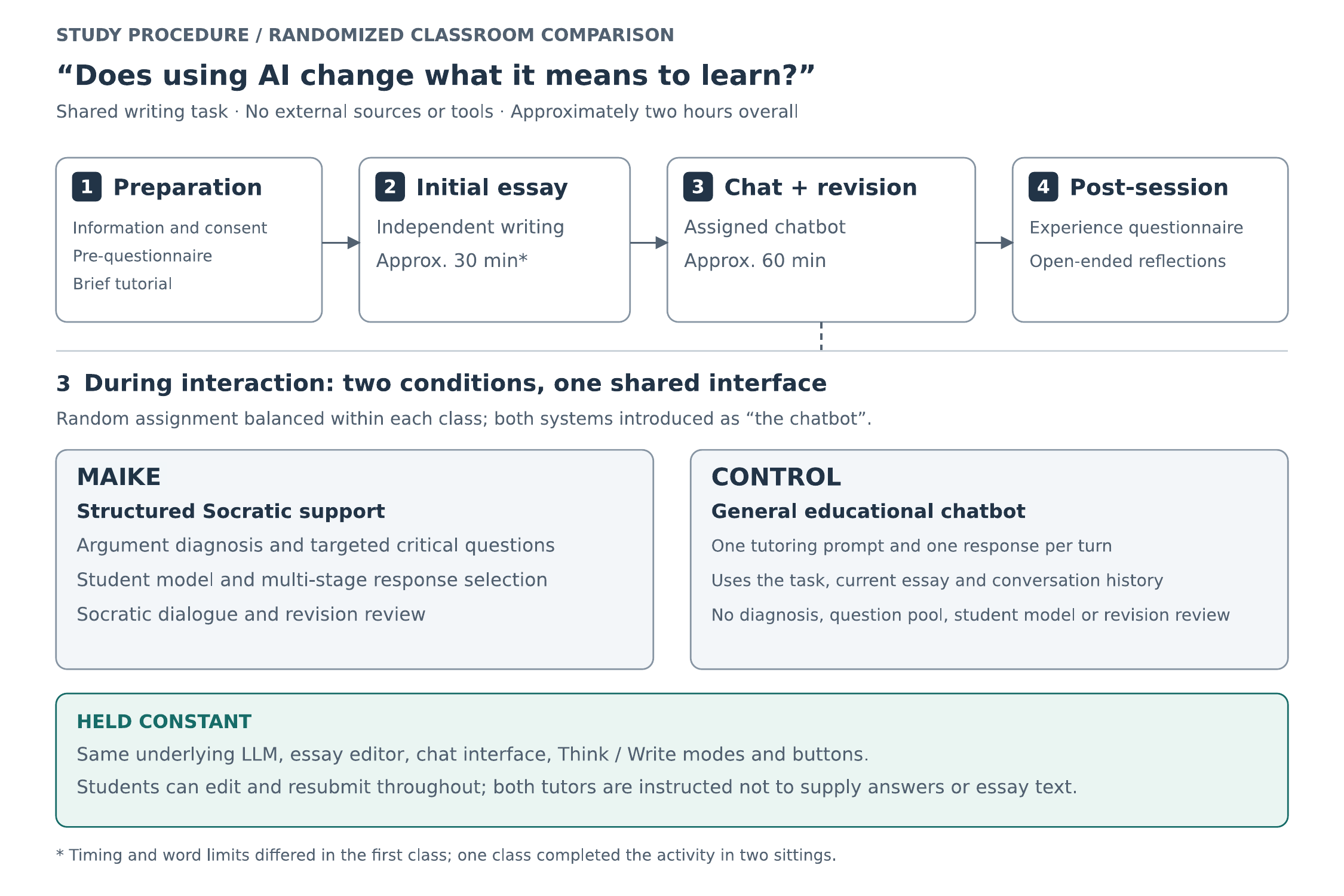}
    \caption{Study procedure and experimental conditions. Students completed a pre-questionnaire, wrote an initial essay independently, interacted with their assigned chatbot while revising, and completed a post-questionnaire. Random assignment was balanced within each class.}
    \Description{Timeline showing preparation, approximately 30 minutes of independent essay writing, 60 minutes of chatbot interaction and revision, and a post-session questionnaire. Below, MAIKE and control are compared, with random assignment within classes and a shared language model and interface.}
    \label{fig:study-procedure}
\end{figure*}

We used a randomized, between-subjects design with two conditions: \Maike and a control chatbot. Participants were randomly assigned to one of the two conditions, with randomization balanced separately within each class cohort. Figure~\ref{fig:study-procedure} summarizes the overall study procedure and the main similarities and differences between the two conditions.

The two conditions were designed to differ primarily in how the chatbot supported the student. Both systems were presented through the same interface, used the same underlying large language model, the same essay editor and submission flow, the same \emph{Think} and \emph{Write} modes, and the same buttons. In both conditions, the chatbot was instructed to provide educational support without writing or rewriting the students' essays or directly supplying text that could serve as the students' answers. The difference between conditions was therefore the interaction strategy and supporting architecture used to deliver this assistance as described in Figure~\ref{fig:study-procedure} panel 3.

In the \textbf{\Maike condition}, the chatbot implemented the structured Socratic architecture described in Section \ref{sec:maike}. \Maike analyzed the student's argument structure and quality, selected targeted critical questions, maintained a student model across the interaction, and evaluated successive essay revisions. These mechanisms were designed to encourage students to articulate, examine, justify and refine their own ideas before translating them into revisions.

In the \textbf{control condition}, students interacted with a general educational chatbot driven by a single tutoring prompt without \Maike's structured components (argument mining, trait scoring, critical-question pool and revision review). On each turn, the control chatbot received the essay question, the current essay, and the conversation history in a single model call, and returned one short message. Like \Maike it was instructed not to provide answers, rewrite the students' essay, or supply corrected text. It therefore provided an active comparison for general conversational and educational support. The complete control prompt is reproduced in the supplementary material, together with the \Maike prompts.

Participants were not informed which condition constituted the experimental intervention or that two different chatbot approaches were being compared. Both systems were introduced simply as ``the chatbot''. This was intended to minimize differences in expectations between conditions and isolate the contribution of \Maike's structured Socratic interaction strategy.

\subsection{Task and procedure}
We next describe each stage of the procedure shown in Figure~\ref{fig:study-procedure}.
Students were asked to write an argumentative essay in response to the ToK-style question, “\textit{Does using AI change what it means to learn?}”. The question was selected because it requires the students to develop and support a position on an open-ended issue, making it well suited to the Theory of Knowledge class.

Students completed the task without access to external sources or tools. They were not permitted to consult reference materials, statistics, citations, search engines, other AI systems or other external sources of information. The assigned chatbot was therefore the only source of interactive support available during the study. The same question, task constraints, and restrictions were applied to both conditions. 
The study was conducted as a classroom activity and followed the same procedure in both conditions and all four classrooms. After a brief introduction of the task, each student received a printed card containing the study URL, a unique participant code, a four-word passphrase and the essay question. 

Students logged into the study platform individually and were first presented with the study information and consent agreement. Students who did not consent to the use of their data completed the classroom activity, yet their data were removed before analysis. The interface was designed so that individual consent decisions were not visible to teachers or to other students. 

\begin{figure*}[t]
    \centering
    \includegraphics[width=\textwidth]{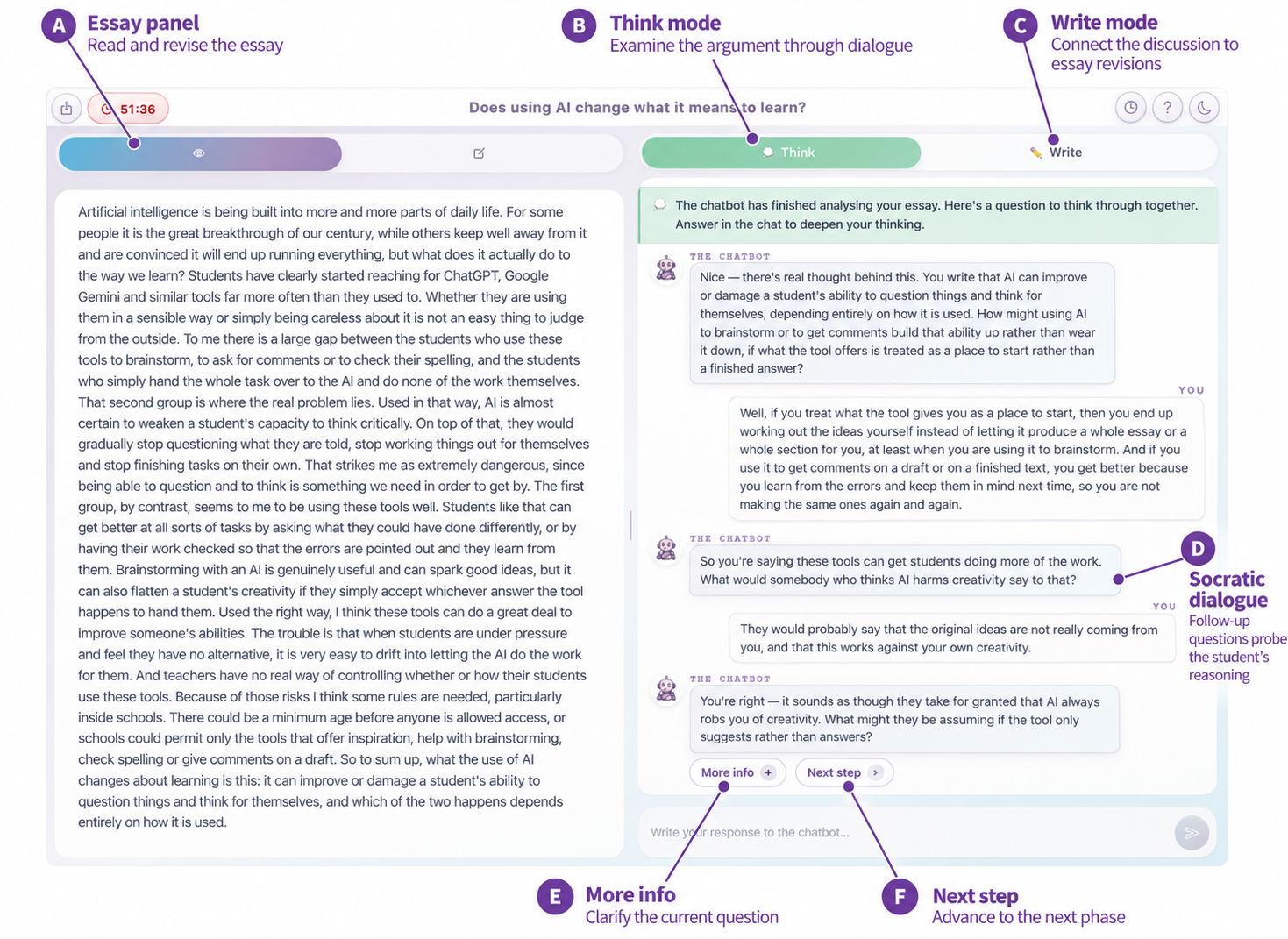}
    \caption{\Maike interface during a Think-mode exchange.
    The essay remains visible alongside the dialogue, where the tutor asks follow-up questions about the student's reasoning.}
    \Description{Annotated interface screenshot with the essay on the left and a Think-mode conversation on the right. Labels identify the essay pane, Think and Write modes, Socratic follow-up questions, the More info clarification button, and the Next step button for advancing.}
    \label{fig:maike-interface}
\end{figure*}

Students then completed the pre-test questionnaires (see supplementary material). After a brief tutorial introducing the platform and task, they wrote their initial essay independently. Once they considered their essay was ready, they began an interaction with the chatbot assigned to their condition. During this phase, students could revise and resubmit their essay as they interacted with the chatbot. In the \Maike condition, each resubmission triggered a new analysis of the essay and generation of a corresponding pool of questions (see Section~\ref{sec:maike-questions}). In the control condition, resubmission initiated another cycle of conversation with the chatbot. 
Figure~\ref{fig:maike-interface} shows the interface. Students could read and revise their text in the essay pane (A) at any time. The dialogue has two modes: \emph{Think} (B) supports reflection on the argument through Socratic questions (D), while \emph{Write} (C) guides students in translating this reflection into essay revisions. Students could request clarification through \emph{More info} (E), logged as \emph{elaborate} or advance through \emph{Next step} (F). We recorded the latter action as \emph{ready to write} to move from \emph{Think} to \emph{Write} mode and \emph{ready to think} when closing the current dialogue round and returning to \emph{Think} mode. In the remainder of the paper, we use these corresponding state names (\emph{elaborate} and \emph{ready to think}).

Across both conditions, students were instructed to use the chatbot as a thinking partner rather than defer to its suggestions, and to retain responsibility for deciding what and when to write. These instructions were identical across conditions. The intervention therefore positioned the chatbot as a source of support while leaving authorship and decision-making to the student. 

The schedule allocated approximately 30 minutes to initial writing and 60 minutes to chatbot interaction and revision, with a dedicated override key letting students advance early once a phase felt complete. Timing, word limits, and session structure varied slightly across the four classes (deviations are in the supplementary material). At the end of the session, students completed a post-test questionnaire, including free-text questions about their experience with the chatbot (see supplementary material). The entire study lasted approximately two hours.

\subsection{Data collection, preprocessing and selection}\label{sec:data}

The application recorded (1) every essay submission, (2) typed chat messages, (3) button presses, (4) keystroke events, and (5) questionnaire responses. Complete essay text was saved at each submission and in periodic snapshots. Keystroke events were used to identify editing episodes and measure edit sizes, rather than to reconstruct essay text.

We used the last available snapshot as the final essay. For 56 of 65 students, it was identical to the last submitted version. For eight, it captured additional edits made after submission. For the remaining student, whose telemetry ended early, we used the last submitted version instead.

For duration-based process measures, each turn was assigned the interval since the immediately preceding logged event. Gaps exceeding 30 minutes were excluded as breaks, including the break between the two sittings of one cohort. Remaining intervals were capped at 10 minutes. These durations include tutor response latency: the median latency per tutor response was 18.3\,s in \Maike and 4.1\,s in control, a consequence of \Maike's multi-stage pipeline, so duration-based measures partly reflect system speed rather than student behavior.

Of 67 students who generated data, 65 consented to research use. Three of these students (one \Maike, two control) barely engaged with the task itself: most of their conversational turns were off-topic, attempts to provoke and hack the chatbot, or requests that it write the essay for them. Their transcripts record almost no task-related interaction to analyze, so the process and essay analyses reported below exclude them, leaving 30 \Maike and 32 control students. 

\section{Analyses and findings}
\label{sec:findings}

We organize the analyses around the three research questions. We first examine how students interacted with each chatbot and moved between dialogue and editing (RQ1). We then assess changes in argumentative quality between their initial and final essays (RQ2), before exploring how their interaction patterns are associated with these changes (RQ3). For each research question, we describe the measures and analysis before presenting the findings. 

\subsection{RQ1: Writing processes}

\subsubsection{Interaction activity and behavioral actions}
\label{sec:beha_states}

\paragraph{Measure construction.} 
\label{sec:states}
To examine how students progressed through the writing task, we represented each session as a sequence of behavioral states extracted deterministically from the interaction logs and the editing telemetry. Each state corresponded to an observable action in the interface, with no human judgment and no time thresholds involved. We distinguish seven states:

\begin{itemize}
  \item \textbf{Think.} The student sent a typed message while the interface was in \emph{Think} mode.
  \item \textbf{Elaborate.} The student pressed the ``Could you elaborate?'' button.
  \item \textbf{Write.} The student sent a typed message while the interface was in \emph{Write} mode.
  \item \textbf{Ready to write.} In Think mode, the student pressed the ``I am ready to write'' button.
  \item \textbf{Edit.} All essay edits captured by the writing telemetry between one chat act (a typed message, a button press, or an essay submission) and the next.
\item \textbf{Ready to think.} The student pressed a button that closed the current round of dialogue and set the interface to a new \emph{Think} mode.
  \item \textbf{Submit.} The student submitted a version of the essay, which also closed the round.
\end{itemize}

\paragraph{Interaction volume.}

Table~\ref{tab:descriptives} summarises the volume of interaction in each condition. Since each condition differs in size ($n_C = 32$ control, $n_S = 30$  \Maike), the raw totals favor the larger control condition. We then report, alongside the totals, the per-participant median and an \emph{adjusted share}, defined as the normalization of the per-participant rates:
\begin{equation}
  \label{eq:adjshare}
  \mathit{Adj. share}_{C} \;=\;
  \frac{T_C / n_C}{\,T_C / n_C + T_S / n_S\,},
\end{equation}
where $T$ is a condition's (indicated by the subscript $C$ or $S$ to refer to control and \Maike, respectively) total count and $n$ its number of participants.

\begin{table}[t]
  \caption{Interaction volume per condition on the primary sample: control (C, $n=32$) and \Maike (S, $n=30$). Medians are per participant within each arm; the adjusted share is defined in Eq.~\ref{eq:adjshare}.}
  \label{tab:descriptives}
  \small
  \setlength{\tabcolsep}{4pt}
  \begin{tabular}{@{}lrrrr@{}}
    \toprule
    & & \multicolumn{3}{c}{C | S} \\
    \cmidrule(l){3-5}
    Measure & Total & Total & Median & Adj.\ share \\
    \midrule
    Student messages (all) &   962 & 502 | 460 & 12.0 | 14.0 & 50.6 | 49.4 \\
    \quad typed messages   &   660 & 344 | 316 &  9.0 | 10.0 & 50.5 | 49.5 \\
    \quad button presses   &   302 & 158 | 144 &  3.5 | 3.5  & 50.7 | 49.3 \\
    Tutor messages         & 1,193 & 637 | 556 & 15.5 | 16.5 & 51.8 | 48.2 \\
    Essay submissions      &   171 & 102 | 69  &  3.0 | 2.0  & 58.1 | 41.9 \\
    \bottomrule
  \end{tabular}
  \Description{Interaction volume per condition on the primary sample (control n=32, Maike n=30): totals, per-condition totals, per-participant medians, and an adjusted share for all student messages, typed messages, button presses, tutor messages, and essay submissions.}
\end{table}

Message volume was descriptively similar across conditions after accounting for group size: adjusted shares were close to 50\% for student messages, button presses, and tutor messages. Students in the control condition submitted their essays more frequently, with a median of three submissions compared with two in \Maike. 

Per student, none of the six volume measures differed by condition: essay submissions were $0.9$ fewer in \Maike (95\% CI $[-1.9, +0.1]$, $p=.08$; Cliff's $\delta=-.21$), the interaction span $28$ minutes longer ($[-20, +75]$, $p=.25$), and the four message counts within $1.5$ messages of equality (all $p \ge .59$).

\paragraph{Behavioral findings.}

\begin{figure}[h]
  \centering
  \includegraphics[width=0.45\textwidth]{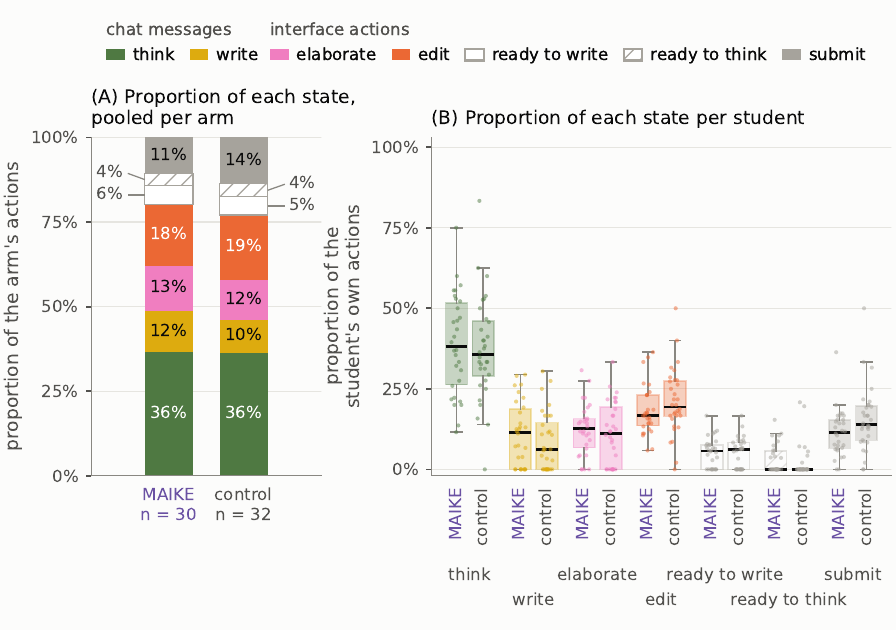}
  \caption{Behavioral states during the session, by condition (\Maike: $n=30$ students, 647 actions; control: $n=32$, 747 actions). (A)~The proportion of each of the seven deterministic behavioral states pooled per condition: two chat-message states (\emph{think}, \emph{write}) and five interface actions (\emph{elaborate}, \emph{edit}, \emph{ready to write}, \emph{ready to think}, \emph{submit}). (B)~The same proportions but per student: each student's count of a state divided by their own total number of actions. Boxes show the median and interquartile range; each dot is one student, \Maike on the left and control on the right of every pair.}
  \Description{Left: stacked bars compare pooled proportions of seven behavioral states; right: paired box plots and dots show proportions for individual students.}
  \label{fig:behavioral-states}
\end{figure}

Figure~\ref{fig:behavioral-states} shows the proportion of each  observable action state described in Section~\ref{sec:states}, pooled per condition~(A) and per student~(B).  The two conditions show a similar composition, with \emph{Think} messages the largest proportion of actions in both (36\% each). 

The clearest differences are \emph{Submit}, with a larger proportion for control actions than for \Maike actions (14\% vs. 11\%), and \emph{Write} messages, with a slightly larger proportion in \Maike (12\% vs. 10\%).
Per student, the submit proportion was $4.1$ percentage points lower in \Maike (95\% CI $[-8.4, +0.1]$, $p=.053$; Cliff's $\delta=-.28$) and the  \emph{Write} messages proportion $2.8$ points higher ($[-1.9, +7.5]$, $p=.25$). No state differed after Holm correction within the seven states.


\subsubsection{Content of the conversational turns and essay edits}
\label{sec:content}

\paragraph{Measure and annotation.}

The observable action states describe what a student did in the interface. They do not describe what a typed message was about. To capture the latter, every typed student turn was assigned to one of four study-specific turn-content categories using  a fixed codebook (see supplementary material). One rater coded the turns and a second rater checked a subset, both working blind to the condition. The categories that describe the content expressed in each turn are the following:

\begin{itemize}
  \item \textbf{Off-task (0).} The student is not doing the task: off-task talk, testing or mocking the system, or asking it to do the student's work (e.g.\ ``write a poem'', ``modify my text'').
  \item \textbf{Confused (1).} The student expresses confusion about what the tutor said or what to do (e.g.\ ``I don't get it'').
  \item \textbf{Surface (2).} On-task interaction about the writing surface or the mechanics of the exchange: requests for feedback, ideas, placement, grammar, translation or synonyms, and statements about what the essay already contains.
  \item \textbf{Idea-level (3).} The turn engages the argument itself (a claim, a reason, an example, a counter-point or a qualification) whether offered by the student or taken up from the tutor.
\end{itemize}

Turns with insufficient content to assign a category were coded \emph{Na} and excluded from the analysis: bare acknowledgments (``done'', ``I did it'') and typing accidents.

The coding follows ICAP's broad principle of interpreting engagement through observable behavior \cite{chi2014icap}, but our categories do not map directly onto its four engagement modes.

 Note that \emph{Idea-level} includes both ideas offered by the student and engagement with ideas supplied in the preceding exchange. Therefore, it should not be interpreted as a direct measure of independent idea generation. 

\paragraph{Edits annotation.}

One annotator coded all eligible episodes using a written codebook (see supplementary material) fixed before coding began. Each episode was shown in context on the coded conversation pages, with condition and student information concealed. . 
An edit episode was classified as substantive when it changed the essay's ideas, claims, examples, or connections between them, rather than only correcting wording, grammar, or formatting. This classification describes what changed, not whether the edit improved the essay.

Of the $261$ edit visits in the primary sample, $33$ were non-substantive ($12.6\%$): $19$ of $118$ in \Maike and $14$ of $143$ in control. This left $228$ substantive edits ($99$ in \Maike and $129$ in control), with a median of $3$ per student in both conditions.

\paragraph{Annotation reliability.}

One author coded all 717 turns and 270 edit episodes with condition and identity concealed. Reliability was checked two ways: a same-coder re-annotation seven days later, and an independent second annotator, blind to condition and to the first coder's labels, coding the first 20 of the 65 conversations. Both checks showed substantial-to-near-perfect agreement (turn-level QWK .75--.80; edit-level exact/raw agreement .92--1.0). Full confidence intervals and the per-check breakdown are reported in the supplementary material.

\paragraph{Results}

Figure~\ref{fig:cognitive-states} shows the proportion of typed student turns in each turn-content category, excluding turns coded \emph{Na}, pooled per condition~(A) and per student~(B).  
On the primary sample, $19$ of the $660$ typed turns were \emph{Na} ($2.9\%$; $6$ of $316$ in \Maike, $13$ of $344$ in control). This left $641$ rated turns, $310$ in \Maike and $331$ in control, with  medians of $10$ and $9$ rated turns per student respectively.

\begin{figure}[h]
  \centering
  \includegraphics[width=0.45\textwidth]{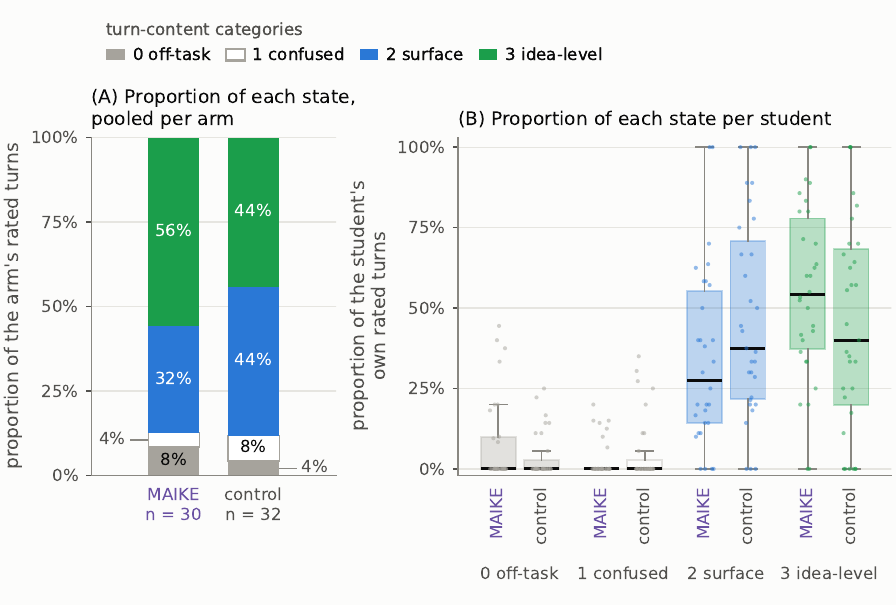}
  \caption{Turn-content categories in students' typed messages, by condition (\Maike: $n=30$ students, 310 coded turns; control: $n=32$ students, 331 coded turns). (A)~The proportion of turns in each category among all coded turns within each condition: \emph{off-task}, \emph{confused}, \emph{surface-level}, and \emph{idea-level}. (B)~The same proportions were calculated separately for each student, by dividing their number of turns in each category by their total number of coded turns. Turns coded \emph{Na} are excluded from both panels.}
  \Description{Stacked bars and per-student box plots compare off-task, confused, surface-level, and idea-level turns.}
  \label{fig:cognitive-states}
\end{figure}

\emph{Idea-level} turns represent the largest proportion of coded turns in \Maike ($56\%$ vs.\ $44\%$ in control), while \emph{surface-level} turns are more prevalent in control ($44\%$ vs.\ $32\%$); the two smaller states run in opposite directions (\emph{off-task}: $8\%$ vs.\ $4\%$ in \Maike; \emph{confused}: $8\%$ vs.\ $4\%$ in control). Panel~(B) shows a wide per-student spread in both conditions, with per-student medians following the same pattern ($54\%$ vs.\ $40\%$ for idea-level, $28\%$ vs.\ $38\%$ for surface). These are descriptive proportions and no test is implied.

The idea-level versus surface-level contrast was tested as the share of idea-level turns among each student's engaged turns (idea-level plus surface-level): per-student means were $.62$ in \Maike and $.48$ in control, a difference of $+15.7$ percentage points (95\% CI $[+1.8, +29.5]$; fractional logit adjusted for class cohort, raw $p=.029$, Holm-adjusted $p=.087$ within the three share outcomes; Cliff's $\delta=+.25$). 

Each student message belongs to either \emph{Think} or \emph{Write} mode. Most of them were sent in \emph{Think} mode in both conditions (\Maike: 234 of 310; control: 268 of 331). In this mode, \emph{idea-level} reasoning was the most frequent state, especially for the \Maike condition ($64\%$ in \Maike and $50\%$ in control of mean rated turns per student), while \emph{surface-level} responses accounted for $25\%$ and $42\%$, respectively. In \emph{Write} mode, this pattern reversed: \emph{surface-level} responses were most frequent ($58\%$ in \Maike, $69\%$ in control), while \emph{idea-level} reasoning decreased to $31\%$ and $17\%$, respectively. \emph{Off-task} and \emph{confused} states remained uncommon across both modes and conditions.

This pattern matches our expectation for the two modes: \emph{Think} mode centers on discussing ideas and arguments, consistent with more \emph{idea-level} turns, while \emph{Write} mode centers on revision mechanics, consistent with more \emph{surface-level} turns.

\subsubsection{Transitions between dialogue and editing}
\label{sec:dynamics}

The two preceding sections describe what students did and what their messages were about, but each counts acts in isolation. Therefore they do not capture how these behaviors unfolded over time. To capture this progression, we combined the message-content coding and interface actions into one chronological sequence for each student. Each act was represented as a state, and each pair of consecutive acts as a transition. This allows us to examine, for example, whether an \emph{idea-level} message is followed by further discussion, an essay edit, or the end of the current dialogue round.

Combining the four message-content categories with the two dialogue modes (\emph{Think} and \emph{Write}) gives eight message states. Adding the five interface actions gives thirteen states in total. Because this produces many possible transitions with only a few observations in each, we reduced them to six main states: \emph{confused}, \emph{surface-level}, \emph{idea-level}, \emph{elaborate}, \emph{edit}, and \emph{move on}. These six retained states are those that mark a distinct next move a student could make: being confused, working at the surface, working on ideas, elaborating, editing, or closing the round.  The distinctions we dropped concern where a turn occurred rather than what it did.

We simplified the sequences in three ways.  First, since the relationship between message content and \emph{Think}/\emph{Write} mode is already examined in Section~\ref{sec:content}, here we focus on what students did next, regardless of the mode in which the message occurred. We therefore combined the \emph{Think} and \emph{Write} versions of each turn-content category. Second, we removed \emph{off-task} turns and the \emph{ready to write} action, which do not represent the substantive progression examined here, and connected the surrounding retained states directly. Third, we combined \emph{ready to think} and \emph{submit} into the label \emph{move on}, because both mark the end of the current dialogue round.

Transitions therefore connect consecutive retained states, which may have been separated by excluded actions in the original log. They describe the order of activities.

The resulting sequences retain codable on-task messages and human-annotated substantive edits. Therefore counts differ from Section~\ref{sec:beha_states}, which includes all messages and edit visits.

\paragraph{Findings.} 

\begin{figure}[h]
  \centering
  \includegraphics[width=\linewidth]{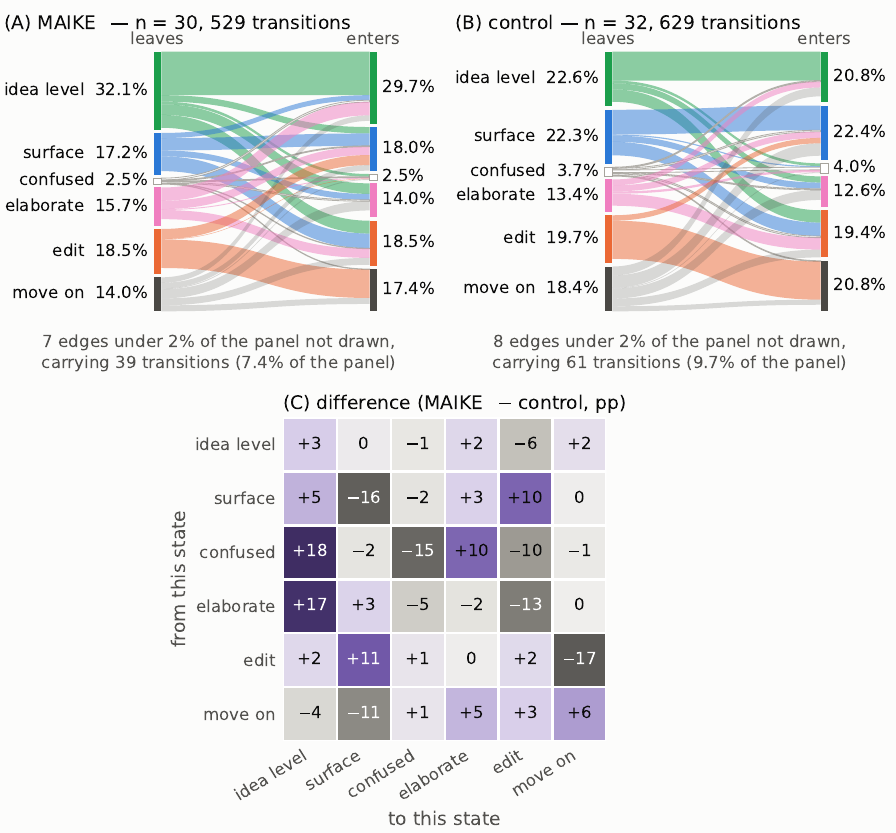}
  \caption{Transitions between the six merged states by condition (\Maike: $n=30$ students, $529$ transitions; control: $n=32$, $629$). Panels~(A) and~(B) show the transition flows within each condition. Ribbon thickness represents the proportion of transitions from the state being left to the next state. Percentages beside each block show the proportion of all transitions leaving (left) and entering (right) that state. Ribbons representing less than $2\%$ of a panel are omitted, except those involving \emph{confused}. Panel~(C) shows the difference in transition proportions between conditions (\Maike minus control, in percentage points); rows indicate the state being left and columns the next state.}
  \Description{Two flow diagrams show transitions among idea-level, surface-level, confused, elaborate, edit, and move-on states. Wider ribbons indicate more frequent transitions. A matrix shows MAIKE-minus-control differences, with previous states in rows and next states in columns.}
  \label{fig:transitions}
\end{figure}

Figure~\ref{fig:transitions} shows the transitions between the six states, one panel per condition, with the difference between them (\Maike minus control) in panel~(C).

Consistent with Figure~\ref{fig:cognitive-states}, \emph{idea-level} messages outnumber \emph{surface-level} messages in the \Maike conditions whereas these two states have a similar presence in the control condition. Furthermore, students who interacted with the control chatbot carried out edits and submits more frequently.

At a high level, the main difference between \Maike and the control  was not simply how often students interacted with the chatbot or edited their essays, but how they moved between dialogue and editing. The sequences of interaction in the control condition progressed more often toward an edit and then to the end of a dialogue round than the sequences of interaction in the \Maike condition where students returned to the dialogue around these edits.

Three states showed the clearest divergence between conditions. \emph{Idea-level} turns were the most self-sustaining state in both conditions (followed by another idea-level turn $56\%$ of the time in \Maike, $53\%$ in control), but students leaving that state diverged: students in the control group more often moved to an essay edit ($23\%$ vs.\ $17\%$) while \Maike students more often requested elaboration ($14\%$ vs.\ $11\%$). \emph{Surface-level} exchanges were markedly less persistent in the \Maike condition (remaining at the surface $30\%$ of the time, vs.\ $46\%$ in control), instead leading more often to an edit or to an idea-level turn. The largest divergence followed an \emph{elaborate} click: students in the \Maike group most often returned to idea-level discussion ($35\%$ vs.\ $18\%$ in control), while control students most often moved directly to editing ($37\%$ vs.\ $24\%$). After an \emph{edit}, sequences in the control group more often ended the round outright ($80\%$ vs.\ $63\%$), while sequences in the \Maike condition more often continued with a further, typically surface-level, exchange ($22\%$ vs.\ $11\%$). An informal manual review suggested these post-edit exchanges sometimes, but not always, followed up on the idea just discussed. \emph{Confused} turns followed a similar directional pattern but rest on few observations ($13$ transitions in \Maike, $23$ in control) and should be read only as a direction, not a stable difference. Full per-state transition percentages, and the supporting manual-review notes, are reported in the supplementary material.

These descriptive patterns suggest different interaction flows depending on the type of chatbot that the students were interacting with. In the control condition, students more often followed a relatively direct progression from dialogue to editing and then moved on, whereas students in the \Maike condition more often moved back and forth between dialogue and the essay. This iterative pattern is particularly visible around requests for elaboration and after edits.

\paragraph{RQ1. How do students' interactions differ between \Maike's diagnostic Socratic approach and general educational chatbot support, particularly in how students move between idea-level discussion and essay revision?}
Returning to RQ1, compared to the control condition, students using \Maike showed a larger descriptive share of \emph{idea-level} turns, while \emph{surface-level} turns were less likely to remain at the surface level discussion, requests for elaboration more often returned students to idea-level reasoning, and edits were more often followed by another exchange with the chatbot. These patterns suggest a more iterative interaction process in the \Maike condition, with students moving back and forth between developing ideas and integrating them into the essay. Conversely, students in the control condition more often moved from discussion to an edit and then proceeded to the next step. 

\subsection{RQ2: Changes in argumentative writing}

\subsubsection{Essay scoring and measurement reliability}
\label{sec:scoring-reliability}

\paragraph{Essay scoring.}
To address RQ2, we assessed changes in the students' argumentative writing between their initial and final essays. 

Two school teachers scored the essays independently, with no access to the dialogue or to information about the student or experimental condition. Essays were presented in a separately randomized order for each teacher, without identifying them as initial or final drafts. Both drafts from a given student were assigned to the same teacher to reduce the influence of stable differences in teacher severity on within-student change.

Each text was scored using five three-band rubrics (0 $=$ weak, 1 $=$ fair, 2 $=$ good), covering three dimensions of argumentative quality adapted from~\citet{wachsmuth-etal-2017-computational} and two from~\citet{wang2026question}. The five traits were logical cogency, rhetorical effectiveness, dialectical reasonableness, evidence, and rebuttal. The complete rubric definitions and scoring guidelines are provided in supplementary material.
To estimate inter-rater reliability, both teachers independently scored twenty essays from ten students (five per condition), sampled to span essay length, overall grade, and prior AI use.

\paragraph{Reliability and trait selection.}

We retained three traits for the outcome analysis---\emph{logical cogency}, \emph{rhetorical effectiveness}, and \emph{rebuttal}--- and excluded dialectical reasonableness and evidence because their observed agreement did not exceed chance ($\kappa \le 0$). Agreement on the retained traits and their composite (\emph{COMP}) ranged from fair to good (QWK .22--.55), with all disagreements limited to adjacent rubric bands (full per-trait statistics are in the supplementary material). Because the same teacher scored both essays for a given student, stable differences in teacher severity affect both measurements and largely cancel out when estimating the pre-to-post change we analyze. Consistent with this, no rubric showed improvement according to one teacher and worsening according to the other.

\subsubsection{Essay changes and condition comparison}\label{sec:essay-change}
\paragraph{Outcome construction and analysis.}

We compared each student's initial and final essay, with the final text defined in  Section~\ref{sec:data}. For the ten double-scored students, we averaged the two teachers' ratings for each trait in each draft before calculating COMP. For all other students, we used the assigned teacher's scores. For each trait and COMP, change was calculated as the final score minus the initial score. Positive values indicate improvement, negative values indicate worsening, and zero indicates no change in score.

\paragraph{Findings.}

\begin{figure}[h]
  \centering
  \includegraphics[width=0.9\linewidth]{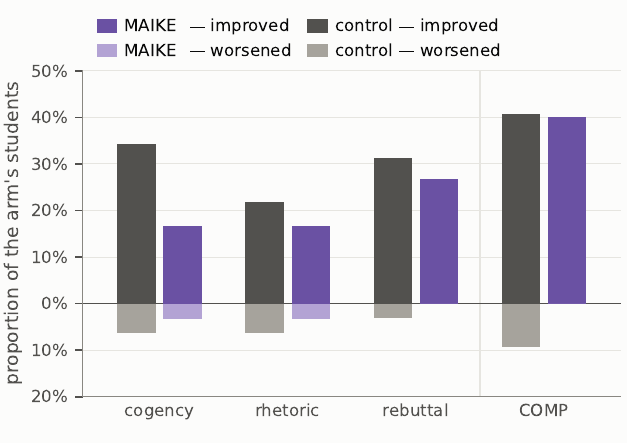}
  \caption{Proportion of students in each condition whose scores improved (above the axis) or worsened (below the axis) between the initial and final essays, for the three retained traits and their composite (COMP: the sum of the three trait scores).}
  \Description{Grouped bars show the proportions of students whose cogency, rhetorical effectiveness, rebuttal, and composite scores improved above zero or worsened below zero. Composite improvement is approximately 40 percent in both conditions.}
  \label{fig:rq1-change}
\end{figure}

Figure~\ref{fig:rq1-change} shows the proportion of students whose scores  improved or worsened between the initial and final essays, for each trait and the composite.

For most students, trait scores remained unchanged between the initial and final essays: approximately 60 to 80\% stayed at the same score. When scores did change, they were usually higher in the final essay.

The proportion of students whose composite score improved was nearly identical across conditions: 12 of 30 \Maike students (40\%) and 13 of 32 control students (41\%). No student in the \Maike condition obtained a composite decline, compared with three students in the control condition (9\%).

At the trait level, the essays produced in the control condition showed both the largest improvement and the largest decline. In the case of \emph{cogency}, the difference between \Maike and the control condition was the largest: 11 of 32 students in the control condition (34\%) improved the cogency of their essays compared with 5 of 30 students in the \Maike condition (17\%). Conversely, 2 students in the control condition (6\%) worsened the cogency of their essays when compared with only 1 student in the \Maike condition (3\%).

These patterns reflect the distinction between improvement on individual traits and net improvement across traits. Among students whose COMP increased, the 13 students in the control condition accounted for 27 upward trait-level changes (2.1 traits per student), compared with 16 among the 12 students in the \Maike condition (1.3 traits per student). Thus, a similar proportion of students improved overall, but gains among the students who improved in the \Maike condition were concentrated in fewer traits.

Declines in individual traits could also be offset by gains elsewhere. The two students in the \Maike condition whose score decreased on one trait each gained at least as much on another trait. Their composite scores therefore remained unchanged or improved.

In sum, similar proportions of students improved their composite scores in both conditions, although improvement on individual traits was less frequent in the \Maike condition. At the same time, no student in the \Maike condition showed a composite decline, compared with three students in the control condition.

\paragraph{RQ2. How does students' argumentative writing change from their initial to final essays, and how do these changes compare across conditions?}
Regarding RQ2, most trait scores remained unchanged between the initial and final essays. Around 40\% of students in each condition improved on the composite score (COMP), with no students in the \Maike condition showing a decline in COMP.

\subsection{RQ3: Associations between process and essay change}
\subsubsection{Process measures and association analysis}\label{sec:associations}
To tackle RQ3, we examine whether characteristics of the student's interaction behavior are associated with the change in essay quality between the initial and final essay and whether these associations differ between conditions.

 We explore thirteen participant-level process measures (Table~\ref{tab:rq3-screen}) derived from the event sequences and annotations described for RQ1. They cover interaction counts, content proportions, engagement drift, time allocation, and temporal pairings between idea-level turns and human-annotated substantive edits. The outcome is the change in COMP defined in Section~\ref{sec:essay-change}.

In particular, two of these measures capture how \emph{idea-level} discussion and substantive editing (as defined in Section~\ref{sec:content}) were ordered during the session. The first, \emph{idea--edit pairs}, counts for each student the number of times an \emph{idea-level} message was followed by substantive edits. Each edit is paired with the most recent \emph{idea-level} turn. Several \emph{idea-level} turns followed by one edit therefore count as one pair. Other messages or interface actions may occur between the turn and the edit, so the pair need not be consecutive. The second measure records whether a student wrote any \emph{idea-level} message at any point after their first substantive edit. We excluded students who never made a substantive edit, since the measure has no starting point for them. Students who did edit but wrote no \emph{idea-level} message afterwards were coded \emph{no}, including those who wrote no further rated messages at all.

Both measures capture the order in which discussion and editing occurred. 

For each of the thirteen participant-level measures, we correlate it with the \emph{COMP} score change separately within each condition  (Spearman's $\rho$). We then take the difference between these two within-condition correlations, $\Delta\rho = \rho_{\textrm{\Maike}} - \rho_{\textrm{control}}$. For any measure, we additionally check that $\Delta\rho$ keeps the same sign when it is recomputed within each teacher's students separately.

Since these analyses rely on naturally occurring variation in the students' interaction behavior rather than experimentally assigned levels of engagement, we interpret the resulting relationships as exploratory associations and we do not treat them as evidence of a causal mediation pathway between interaction behavior and writing improvement. 

Supplementary material provides the operational definitions of the process measures, missing-value handling, bootstrap procedure, and teacher-specific sensitivity checks. 

\subsubsection{Association findings}
Table~\ref{tab:rq3-screen} summarizes these associations.
\begin{table}[t]
  \caption{
  Between-condition differences in the associations between process measures and essay-score change. For each of the 13 process measures, we report Spearman's~$\rho$ with the change in composite score ($\Delta$COMP, final minus initial) within each condition. Difference between correlations: $\Delta\rho = \rho_{\textrm{\Maike}} - \rho_{\textrm{control}}$ is reported with a 95\% bootstrap confidence interval, based on 2{,}000 resamples of students within each condition. Rows whose interval excludes zero are in bold.}
  \label{tab:rq3-screen}
  \small
  \setlength{\tabcolsep}{3pt}
  \begin{tabularx}{\columnwidth}{@{}>{\raggedright\arraybackslash}Xrrr@{}}
    \toprule
    Process measure & $\rho_{\textrm{\Maike}}$ & $\rho_{\textrm{control}}$ & $\Delta\rho$ [95\% CI] \\
    \midrule
    \textbf{Idea-level turns followed by a substantive edit (count)} & \textbf{+.49} & \textbf{$-$.14} & \textbf{+.63 [+.21, +1.03]} \\
    \textbf{Returned to idea-level after the first edit (yes/no)} & \textbf{+.39} & \textbf{$-$.15} & \textbf{+.54 [+.09, +.97]} \\
    Engagement drift (late $-$ early) & +.13 & $-$.21 & +.34 [$-$.10, +.73] \\
    Rated conversation turns (count) & +.16 & $-$.17 & +.33 [$-$.20, +.80] \\
    Confused proportion of rated turns & +.03 & +.31 & $-$.28 [$-$.73, +.22] \\
    Elaborate clicks (count) & +.33 & +.12 & +.21 [$-$.36, +.69] \\
    Idea-level turns (count) & +.05 & $-$.15 & +.20 [$-$.30, +.65] \\
    Essay submissions (count) & +.02 & $-$.05 & +.07 [$-$.44, +.56] \\
    Idea-level turns before the first submission & $-$.19 & $-$.13 & $-$.07 [$-$.63, +.46] \\
    Idea-level proportion of engaged turns & $-$.03 & +.02 & $-$.06 [$-$.61, +.46] \\
    Proportion of attributed conversation time at the idea level & $-$.05 & $-$.02 & $-$.04 [$-$.56, +.47] \\
    Idea-level proportion of rated turns & $-$.03 & $-$.01 & $-$.02 [$-$.55, +.50] \\
    Substantive edits (count) & +.08 & +.09 & $-$.01 [$-$.51, +.49] \\
    \bottomrule
  \end{tabularx}
  \Description{For 13 process measures, Spearman's rho with the change in composite essay score within Maike and within control, and the Maike-minus-control difference with a 95\% bootstrap confidence interval; the two rows whose interval excludes zero are in bold.}
\end{table}

For most process measures, we found no clear evidence that their association with essay-score changes differed between conditions. The clearest differences concerned the sequence of \emph{idea-level} turns and substantive edits.
In the \Maike condition, students with more \emph{idea-level} turns followed by substantive edits tended to show greater score gains ($\rho=.49$). Interestingly, the corresponding correlation was weakly negative in the control condition ($\rho=-.14$), with a between-condition difference of $\Delta\rho=.63$ (95\% CI $[.21,1.03]$). Returning to \emph{idea-level} turns after the first significant edit showed a similar between-condition difference ($\Delta\rho=.54$, 95\% CI $[.09,.97]$). By comparison, the association between the number of substantive edits and score gains did not clearly differ between conditions.

\paragraph{RQ3. How is the coordination of idea-level discussion and essay revision associated with changes in the students’
argumentative writing?}
Regarding RQ3, in the \Maike condition, students who more often transitioned from \emph{idea-level} discussion to substantive revision tended to show greater improvement in their argumentative writing. The corresponding association was weaker in the control condition, yielding a similar between-condition pattern for returning to idea-level discussion after editing. These exploratory associations suggest that, in \Maike, iterative movement between reasoning and revision may be particularly relevant to writing improvement. 

The supplementary material contains a summary of the self-reported feedback provided by the students regarding their user experience. We found no evidence of a difference between conditions: across seven post-test scales, every between-condition interval spanned zero (Cliff's $\delta$ from $-.15$ to $+.12$; all Holm-adjusted $p = 1.0$). 

\section{Discussion and implications}
\label{sec:discussion}

Our findings suggest that the contribution of AI support for argumentative writing may be better understood in terms of how students engage with the writing process than in terms of the quality of their final text.  Although approximately the same proportion of students improved their composite writing score in both conditions, students using \Maike interacted with the system differently: they spent a larger share of their interaction at the idea level and more often moved back and forth between discussion and essay revision. Moreover, the exploratory relationship between idea-level discussion, substantive revision, and writing improvement was more positive in \Maike than in the control condition. Given the exploratory nature of our study, these patterns do not establish that \Maike or any particular interaction sequence causes better writing. Instead, they point to a set of interaction-design opportunities around how AI can connect students' reasoning to their own writing activity. 

\subsection{Designing for reasoning--revision loops}
A central implication is that AI writing support can be designed around \emph{reasoning--revision loops} rather than treating reasoning and writing as sequential activities. In the \Maike group, students more often returned to idea-level discussion after editing and other surface-level exchanges, suggesting that ideas remained open to reconsideration as the essay developed. By contrast, the control condition more often followed a progression from discussion to editing and then to the next step. This difference suggests that productive AI support may involve helping students repeatedly move between developing an idea and acting on that idea in the text. This recursive pattern echoes long-standing accounts of writing as a non-linear process in which planning, translating and reviewing interact throughout a task rather than proceeding in a fixed sequence \cite{flower1981cognitive}, and it specifically illustrates Scardamalia and Bereiter's account of reflective, knowledge-transforming composition as a two-way exchange between a content problem space and a rhetorical problem space \cite{scardamalia1984teachability}: our findings suggest that AI dialogue can be designed to sustain this two-way exchange. 

For designers, this means that the unit of interaction need not be the individual chatbot response. Instead, systems can be designed around \emph{interaction trajectories} that connect conversational reasoning with changes in the students' essay. After a student revised a claim, for example, the system could prompt them to reconsider whether the revision affects the underlying reasoning or evidence. Similarly, after identifying a weakness in an argument, the system could direct the student toward revising the relevant part of the essay and subsequently revisit whether the revision resolved the problem \cite{tankelevitch2024metacognitive}. Such mechanisms could make the transition between reasoning and writing an explicit part of the interaction design. 

\subsection{Making reflection consequential}
The relationship we observed between idea-level discussion followed by substantive revision and writing improvement further suggests that reflection may be most valuable when it becomes consequential for the students' essay. Socratic questioning can encourage students to consider assumptions, alternatives, or weaknesses, but reflection alone does not guarantee that these insights will affect the resulting essay. The relevant design question is therefore how to help students translate reflection into decisions about their writing. This is consistent with HCI evidence that how an AI system frames its contribution shapes whether users engage their own reasoning: framing feedback as questions rather than as direct explanations improved people's own logical discernment of flawed statements \cite{danry2023don}, whereas AI that supplies opinions or ready-made content can shift what users themselves come to believe \cite{jakesch2023co}. By diagnosing issues but only ever asking about them, rather than stating or resolving them, \Maike is designed to occupy the first role without risking the second. 

This finding points toward designing AI interventions that connect a student's reasoning to an actionable change in the essay while leaving the actual revision to the student. For example, rather than asking a series of disconnected critical questions, a system could identify an unresolved issue, ask the student to consider its implications, and then prompt them to determine where and how their argument should change. After the revision, the system could return to the earlier issue and invite the student to evaluate whether their change addressed it. The resulting interaction would make reflection part of an ongoing cycle of reasoning and action instead of an isolated conversational activity. 

\subsection{Designing state-aware AI writing support}
The iterative patterns observed in our study also have implications for how conversational writing systems represent and respond to the students' evolving activity. A chatbot that responds primarily to the latest message may treat each exchange as an independent interaction, even though its meaning depends on what the student has previously considered and changed in the essay. In contrast, supporting iterative writing requires some representation of the student's reasoning and writing state: what claims they have developed, which weaknesses remain unresolved, and how their recent revisions relate to earlier discussion. 

\Maike's diagnostic architecture and student model represent one approach to this problem. More generally, AI writing systems could use information about the student's prior reasoning and recent revisions to determine the appropriate form and timing of support \cite{wambsganss2021arguetutor,mouchel2023revision}. The system might prioritize questioning when an argument remains underdeveloped, reconnect a surface-level request to an unresolved idea or prompt reconsideration after a substantive revision. This suggests a shift from designing systems that are primarily response-aware toward systems that are sensitive to the student's evolving position within the broader reasoning--writing process. This framing is aligned with recent HCI work arguing that generative-AI interfaces should explicitly support users' metacognition, i.e., their ability to monitor and regulate their own thinking, rather than optimize only the quality of individual outputs \cite{tankelevitch2024metacognitive}. A state-aware writing tutor operationalizes this idea for argumentative writing: it is the system's model of the student's reasoning and revisions, not just the latest message, that should determine when and how to intervene. 

\subsection{Evaluating AI writing support through interaction trajectories}

The similar writing outcomes, despite different interaction patterns, suggest that evaluating AI writing support requires looking beyond essay quality. An evaluation focused only on initial and final essays would suggest relatively little difference between the systems, despite the substantially different interaction patterns we observed. This finding illustrates how essay scoring measures can obscure meaningful differences in the processes through which students engage with AI and reflects the distinction the ICAP framework makes between passive, active, constructive and interactive forms of engagement, in which the mode of cognitive engagement predicts learning outcomes \cite{chi2014icap}: our two conditions produced comparable volumes of interaction but different proportions of idea-level engagement, and it was this qualitative difference in engagement that our exploratory associations point to as relevant to writing improvement. 

We therefore suggest considering educational AI at three complementary levels: (1) the essays the students produce, (2) the interaction they have with the system and (3) the coordination between interaction and essay change. The third level is particularly relevant for writing support. Thus, rather than asking only whether students discussed ideas or whether their essays improved, researchers can examine whether reasoning was followed by substantive revision, whether revisions prompted renewed reasoning, and which interaction trajectories coincide with changes in writing \cite{flower1981cognitive, matcha2019analytics}. These process-oriented measures can identify promising mechanisms for future hypothesis-driven studies, even when exploratory studies are not powered to establish causal effects, complementing prior human-AI writing research that treats trajectory of an interaction as the primary unit of analysis \cite{lee2022coauthor}. We extend this approach by relating trajectories to the coordination between reasoning and revision. 

Moreover, more interaction is not necessarily better. Our findings suggest that the design target should be \emph{consequential interaction}, \emph{i.e.}, interaction that changes how students reason about, act on, and reconsider their own writing. Future evaluations could move beyond measures of interaction volume toward measures of meaningful transitions between cognitive activity and essay revision actions \cite{matcha2019analytics}. 

The composite-decline asymmetry we observed between conditions offers a tentative illustration of this point: no student in the \Maike condition worsened on the composite score compared to three students (9\%) in the control condition. An edit that is not anchored to a diagnosed weakness can move the essay in either direction, whereas \Maike's diagnosis-gated questioning might act as a guardrail against such unproductive changes. This observation is consistent with reports that the quality of AI feedback does not translate into better revisions \cite{farrokhnia2026generative} and that AI-assisted revision tends to correct surface-level issues more reliably than it improves deeper structure \cite{pretorius2026mediating}. 

\subsection{Supporting student responsibility through AI}
Overall, our findings point to an alternative role for AI in educational writing. Instead of serving as a generator or editor of text, AI can be designed to orchestrate the students' engagement with their own reasoning and writing decisions \cite{wood1976scaffold}. \Maike demonstrates one such design space by providing diagnostic and Socratic support while leaving the production of essay text to the student. The system therefore contributes to the process without taking over the product. This operationalizes the \emph{scaffold, not substitute} principle for argumentative writing \cite{favero2026substitution,wood1976scaffold}. 

This effort is not free, but its cost appears small. Self-reported experience was lower for \Maike than for the control yet every between-condition interval included zero and absolute ratings. The fact that a system that asks students to do more of the reasoning work themselves was not rated meaningfully worse than one that did not is itself a positive result for this design space: preserving epistemic responsibility need not come at a high cost to how the interaction feels to use. 

This distinction is of paramount importance for educational AI where the goal is not necessarily to maximize the amount of writing performed by the system \cite{favero2026substitution}. Instead, AI can be used to structure opportunities for the students to question, reconsider, and act on their own ideas \cite{hugenroth2026critical}. Our findings do not establish that this approach produces superior writing outcomes, but they suggest that preserving student responsibility and supporting reasoning need not require removing AI from the writing process. Designing AI to support the transitions between thinking and writing offers a middle ground between unaided writing and AI-generated composition, complementing recent question-based writing supports \cite{wang2026question,namaziandost2026socratic} and Socratic systems that pair argument analysis with later revision \cite{hugenroth2026critical}: instead of treating reasoning and revision as separate phases linked only by a shared question, our process-level findings suggest that it is the repeating coordination between the two that is worth designing and evaluating for. 

Thus, a useful design approach for AI-supported argumentative writing entails addressing how the interaction with the AI tool connects the students' reasoning to the revisions they make themselves. The reasoning--revision loop identified in our study provides a promising design target for future systems, while the corresponding interaction trajectories provide a basis for evaluating whether such systems meaningfully engage students in their own writing process.

\section{Limitations and future work}

Our work is not without limitations, particularly our exploratory classroom study which evaluates \Maike as a complete system. The active control chatbot shared the same interface and several features with \Maike, including withholding answers and essay text and implementing \emph{Think} and \emph{Write} modes. The classroom study therefore isolates neither Socratic questioning nor the Think–Write split. 
The evidence obtained in our exploratory study is limited by the sample size and measures. Given the small sample size, the analyses are exploratory. 
The study covered only one essay prompt, one 2-hour classroom session, and one population in which English was mostly second language yet regularly used in their education  (Section~\ref{sec:experimental-study}). Most importantly, we measured interaction behavior and changes to one supported essay, not durable learning, transfer, independent critical thinking, learner agency, or reduced reliance on AI.

A further limitation concerns the reliability of \Maike's own diagnostic pipeline. A weak component misclassified as strong, or a trait misjudged,
would propagate into a critical question grounded in that misdiagnosis, and a student has no way in the current interface to tell a well-targeted question from a misdirected one. Students have been shown to reject AI-generated feedback when they believe it misreads their intended meaning, and to do so more often for content-level than for form-level comments \cite{chen2026barriers}. An AI tutor that occasionally misdiagnoses risks the same loss of trust. Future work should examine how students respond when a critical question is visibly misaligned with their essay. 

Future work should also include a larger study varying support for connecting discussion and revision while holding the remaining assistance constant. Combining process traces with analysis of the actual ideas carried into revisions would test the proposed connection more directly. Repeated sessions and subsequent writing without AI would then help establish whether this support contributes to learning beyond the immediate essay.

\section{Conclusion}

We have presented \Maike, an educational chatbot that includes Socratic questioning to support argumentative writing while leaving students responsible for developing their ideas and writing their text. \Maike analyzes the student’s essay to guide its questions, helping them  reason through ideas and connect their reasoning to essay revisions. In an exploratory classroom study with 65 secondary school students, students using \Maike produced a larger proportion of idea-level turns and moved more often between dialogue and editing than control students. These descriptive differences, however, did not translate into a clear advantage in essay quality or students’ reported experience.
Within the \Maike group, students who more often followed idea-level dialogue with substantive edits tended to improve more. Although this association does not establish causality or show that the edits drew directly on the dialogue, it points to the connection between reflection and revision as a potentially important mechanism. Future studies should examine this link directly and test whether such benefits persist in later writing without AI.

\begin{acks}
This work has been partially supported by a nominal grant received at the ELLIS Unit Alicante Foundation from the Regional Government of Valencia in Spain (Resolución of the Generalitat Valenciana, Conselleria de Innovación, Industria, Comercio y Turismo, Dirección General de Innovación). L.F. has also been partially funded by the Bank Sabadell Foundation. This project has received funding from the European Union’s Horizon  Europe  research  and  innovation  programme  under  grant  agreement  101120237 (ELIAS).   
\end{acks}

\begin{thebibliography}{43}

\ifx \showCODEN    \undefined \def \showCODEN     #1{\unskip}     \fi
\ifx \showISBNx    \undefined \def \showISBNx     #1{\unskip}     \fi
\ifx \showISBNxiii \undefined \def \showISBNxiii  #1{\unskip}     \fi
\ifx \showISSN     \undefined \def \showISSN      #1{\unskip}     \fi
\ifx \showLCCN     \undefined \def \showLCCN      #1{\unskip}     \fi
\ifx \shownote     \undefined \def \shownote      #1{#1}          \fi
\ifx \showarticletitle \undefined \def \showarticletitle #1{#1}   \fi
\ifx \showURL      \undefined \def \showURL       {\relax}        \fi
\providecommand\bibfield[2]{#2}
\providecommand\bibinfo[2]{#2}
\providecommand\natexlab[1]{#1}
\providecommand\showeprint[2][]{arXiv:#2}

\bibitem[Brown et~al\mbox{.}(2020)]%
        {brown2020language}
\bibfield{author}{\bibinfo{person}{Tom Brown}, \bibinfo{person}{Benjamin Mann}, \bibinfo{person}{Nick Ryder}, \bibinfo{person}{Melanie Subbiah}, \bibinfo{person}{Jared~D Kaplan}, \bibinfo{person}{Prafulla Dhariwal}, \bibinfo{person}{Arvind Neelakantan}, \bibinfo{person}{Pranav Shyam}, \bibinfo{person}{Girish Sastry}, \bibinfo{person}{Amanda Askell}, \bibinfo{person}{Sandhini Agarwal}, \bibinfo{person}{Ariel Herbert-Voss}, \bibinfo{person}{Gretchen Krueger}, \bibinfo{person}{Tom Henighan}, \bibinfo{person}{Rewon Child}, \bibinfo{person}{Aditya Ramesh}, \bibinfo{person}{Daniel Ziegler}, \bibinfo{person}{Jeffrey Wu}, \bibinfo{person}{Clemens Winter}, \bibinfo{person}{Chris Hesse}, \bibinfo{person}{Mark Chen}, \bibinfo{person}{Eric Sigler}, \bibinfo{person}{Mateusz Litwin}, \bibinfo{person}{Scott Gray}, \bibinfo{person}{Benjamin Chess}, \bibinfo{person}{Jack Clark}, \bibinfo{person}{Christopher Berner}, \bibinfo{person}{Sam McCandlish}, \bibinfo{person}{Alec Radford}, \bibinfo{person}{Ilya Sutskever}, {and}
  \bibinfo{person}{Dario Amodei}.} \bibinfo{year}{2020}\natexlab{}.
\newblock \showarticletitle{Language Models are Few-Shot Learners}. In \bibinfo{booktitle}{\emph{Advances in Neural Information Processing Systems}}, \bibfield{editor}{\bibinfo{person}{H.~Larochelle}, \bibinfo{person}{M.~Ranzato}, \bibinfo{person}{R.~Hadsell}, \bibinfo{person}{M.F. Balcan}, {and} \bibinfo{person}{H.~Lin}} (Eds.), Vol.~\bibinfo{volume}{33}. \bibinfo{publisher}{Curran Associates, Inc.}, \bibinfo{pages}{1877--1901}.
\newblock
\urldef\tempurl%
\url{https://proceedings.neurips.cc/paper_files/paper/2020/file/1457c0d6bfcb4967418bfb8ac142f64a-Paper.pdf}
\showURL{%
\tempurl}


\bibitem[Chen et~al\mbox{.}(2026)]%
        {chen2026barriers}
\bibfield{author}{\bibinfo{person}{Ziqi Chen}, \bibinfo{person}{Xinhua Zhu}, \bibinfo{person}{Qi Lu}, {and} \bibinfo{person}{Wei Wei}.} \bibinfo{year}{2026}\natexlab{}.
\newblock \showarticletitle{{L2} students' barriers in engaging with form and content-focused {AI}-generated feedback in revising their compositions}.
\newblock \bibinfo{journal}{\emph{Computer Assisted Language Learning}} \bibinfo{volume}{39}, \bibinfo{number}{3} (\bibinfo{year}{2026}), \bibinfo{pages}{715--735}.
\newblock
\href{https://doi.org/10.1080/09588221.2024.2422478}{doi:\nolinkurl{10.1080/09588221.2024.2422478}}


\bibitem[Chi and Wylie(2014)]%
        {chi2014icap}
\bibfield{author}{\bibinfo{person}{Michelene~TH Chi} {and} \bibinfo{person}{Ruth Wylie}.} \bibinfo{year}{2014}\natexlab{}.
\newblock \showarticletitle{The ICAP framework: Linking cognitive engagement to active learning outcomes}.
\newblock \bibinfo{journal}{\emph{Educational psychologist}} \bibinfo{volume}{49}, \bibinfo{number}{4} (\bibinfo{year}{2014}), \bibinfo{pages}{219--243}.
\newblock


\bibitem[Danry et~al\mbox{.}(2023)]%
        {danry2023don}
\bibfield{author}{\bibinfo{person}{Valdemar Danry}, \bibinfo{person}{Pat Pataranutaporn}, \bibinfo{person}{Yaoli Mao}, {and} \bibinfo{person}{Pattie Maes}.} \bibinfo{year}{2023}\natexlab{}.
\newblock \showarticletitle{Don’t just tell me, ask me: Ai systems that intelligently frame explanations as questions improve human logical discernment accuracy over causal ai explanations}. In \bibinfo{booktitle}{\emph{Proceedings of the 2023 CHI conference on human factors in computing systems}}. \bibinfo{pages}{1--13}.
\newblock


\bibitem[Dong et~al\mbox{.}(2024)]%
        {dong2024icl}
\bibfield{author}{\bibinfo{person}{Qingxiu Dong}, \bibinfo{person}{Lei Li}, \bibinfo{person}{Damai Dai}, \bibinfo{person}{Ce Zheng}, \bibinfo{person}{Jingyuan Ma}, \bibinfo{person}{Rui Li}, \bibinfo{person}{Heming Xia}, \bibinfo{person}{Jingjing Xu}, \bibinfo{person}{Zhiyong Wu}, \bibinfo{person}{Baobao Chang}, \bibinfo{person}{Xu Sun}, \bibinfo{person}{Lei Li}, {and} \bibinfo{person}{Zhifang Sui}.} \bibinfo{year}{2024}\natexlab{}.
\newblock \showarticletitle{A Survey on In-context Learning}. In \bibinfo{booktitle}{\emph{Proceedings of the 2024 Conference on Empirical Methods in Natural Language Processing}}, \bibfield{editor}{\bibinfo{person}{Yaser Al-Onaizan}, \bibinfo{person}{Mohit Bansal}, {and} \bibinfo{person}{Yun-Nung Chen}} (Eds.). \bibinfo{publisher}{Association for Computational Linguistics}, \bibinfo{address}{Miami, Florida, USA}, \bibinfo{pages}{1107--1128}.
\newblock
\href{https://doi.org/10.18653/v1/2024.emnlp-main.64}{doi:\nolinkurl{10.18653/v1/2024.emnlp-main.64}}


\bibitem[D’Mello and Graesser(2012)]%
        {d2012dynamics}
\bibfield{author}{\bibinfo{person}{Sidney D’Mello} {and} \bibinfo{person}{Art Graesser}.} \bibinfo{year}{2012}\natexlab{}.
\newblock \showarticletitle{Dynamics of affective states during complex learning}.
\newblock \bibinfo{journal}{\emph{Learning and Instruction}} \bibinfo{volume}{22}, \bibinfo{number}{2} (\bibinfo{year}{2012}), \bibinfo{pages}{145--157}.
\newblock


\bibitem[Faigley and Witte(1981)]%
        {faigley1981analyzing}
\bibfield{author}{\bibinfo{person}{Lester Faigley} {and} \bibinfo{person}{Stephen Witte}.} \bibinfo{year}{1981}\natexlab{}.
\newblock \showarticletitle{Analyzing revision}.
\newblock \bibinfo{journal}{\emph{College Composition and Communication}} \bibinfo{volume}{32}, \bibinfo{number}{4} (\bibinfo{year}{1981}), \bibinfo{pages}{400--414}.
\newblock
\href{https://doi.org/10.58680/ccc198115887}{doi:\nolinkurl{10.58680/ccc198115887}}


\bibitem[Farrokhnia et~al\mbox{.}(2026)]%
        {farrokhnia2026generative}
\bibfield{author}{\bibinfo{person}{Mohammadreza Farrokhnia}, \bibinfo{person}{Saeed Latifi}, \bibinfo{person}{Pantelis~M. Papadopoulos}, \bibinfo{person}{Loes Hogenkamp}, \bibinfo{person}{Hannie Gijlers}, \bibinfo{person}{Hassan Khosravi}, {and} \bibinfo{person}{Omid Noroozi}.} \bibinfo{year}{2026}\natexlab{}.
\newblock \showarticletitle{Generative {AI} offers more, but students revise less: Comparing the effects of teacher and {AI} feedback on student essay revisions}.
\newblock \bibinfo{journal}{\emph{International Journal of Educational Technology in Higher Education}} \bibinfo{volume}{23}, \bibinfo{number}{1} (\bibinfo{year}{2026}), \bibinfo{pages}{6}.
\newblock
\href{https://doi.org/10.1186/s41239-026-00579-9}{doi:\nolinkurl{10.1186/s41239-026-00579-9}}


\bibitem[Favero et~al\mbox{.}(2025)]%
        {favero-etal-2025-ellis}
\bibfield{author}{\bibinfo{person}{Lucile Favero}, \bibinfo{person}{Daniel Frases}, \bibinfo{person}{Juan~Antonio P{\'e}rez-Ortiz}, \bibinfo{person}{Tanja K{\"a}ser}, {and} \bibinfo{person}{Nuria Oliver}.} \bibinfo{year}{2025}\natexlab{}.
\newblock \showarticletitle{{ELLIS} {A}licante at {CQ}s-Gen 2025: Winning the critical thinking questions shared task: {LLM}-based question generation and selection}. In \bibinfo{booktitle}{\emph{Proceedings of the 12th Argument mining Workshop}}, \bibfield{editor}{\bibinfo{person}{Elena Chistova}, \bibinfo{person}{Philipp Cimiano}, \bibinfo{person}{Shohreh Haddadan}, \bibinfo{person}{Gabriella Lapesa}, {and} \bibinfo{person}{Ramon Ruiz-Dolz}} (Eds.). \bibinfo{publisher}{Association for Computational Linguistics}, \bibinfo{address}{Vienna, Austria}, \bibinfo{pages}{322--331}.
\newblock
\showISBNx{979-8-89176-258-9}
\href{https://doi.org/10.18653/v1/2025.argmining-1.31}{doi:\nolinkurl{10.18653/v1/2025.argmining-1.31}}


\bibitem[Favero et~al\mbox{.}(2026a)]%
        {favero2026substitution}
\bibfield{author}{\bibinfo{person}{Lucile Favero}, \bibinfo{person}{Juan~Antonio P{\'e}rez-Ortiz}, \bibinfo{person}{Tanja K{\"a}ser}, {and} \bibinfo{person}{Nuria Oliver}.} \bibinfo{year}{2026}\natexlab{a}.
\newblock \showarticletitle{From Substitution to Scaffolding: Breaking the Self-Reinforcing Harm Cycle of AI in Education (and Beyond)}.
\newblock \bibinfo{journal}{\emph{arXiv preprint}} (\bibinfo{year}{2026}).
\newblock


\bibitem[Favero et~al\mbox{.}(2026b)]%
        {favero2026leveraging}
\bibfield{author}{\bibinfo{person}{Lucile Favero}, \bibinfo{person}{Juan~Antonio P{\'e}rez-Ortiz}, \bibinfo{person}{Tanja K{\"a}ser}, {and} \bibinfo{person}{Nuria Oliver}.} \bibinfo{year}{2026}\natexlab{b}.
\newblock \showarticletitle{Leveraging Small {LLMs} for Argument Mining in Education: Argument Component Identification, Classification, and Assessment}. In \bibinfo{booktitle}{\emph{Proceedings of the 25th Workshop on Computational Models of Natural Argument ({CMNA} 2025)}} \emph{(\bibinfo{series}{CEUR Workshop Proceedings}, Vol.~\bibinfo{volume}{4166})}, \bibfield{editor}{\bibinfo{person}{Giulia D'Agostino}, \bibinfo{person}{Floriana Grasso}, \bibinfo{person}{Nancy~L. Green}, \bibinfo{person}{Roos~J. Scheffers}, \bibinfo{person}{Jodi Schneider}, {and} \bibinfo{person}{Simon Wells}} (Eds.). \bibinfo{publisher}{CEUR-WS.org}, \bibinfo{pages}{11--30}.
\newblock
\showISSN{1613-0073}
\urldef\tempurl%
\url{https://ceur-ws.org/Vol-4166/paper2.pdf}
\showURL{%
\tempurl}


\bibitem[Flower and Hayes(1981)]%
        {flower1981cognitive}
\bibfield{author}{\bibinfo{person}{Linda Flower} {and} \bibinfo{person}{John~R. Hayes}.} \bibinfo{year}{1981}\natexlab{}.
\newblock \showarticletitle{A cognitive process theory of writing}.
\newblock \bibinfo{journal}{\emph{College Composition and Communication}} \bibinfo{volume}{32}, \bibinfo{number}{4} (\bibinfo{year}{1981}), \bibinfo{pages}{365--387}.
\newblock
\href{https://doi.org/10.58680/ccc198115885}{doi:\nolinkurl{10.58680/ccc198115885}}


\bibitem[Geng et~al\mbox{.}(2023)]%
        {geng2023gcd}
\bibfield{author}{\bibinfo{person}{Saibo Geng}, \bibinfo{person}{Martin Josifoski}, \bibinfo{person}{Maxime Peyrard}, {and} \bibinfo{person}{Robert West}.} \bibinfo{year}{2023}\natexlab{}.
\newblock \showarticletitle{Grammar-Constrained Decoding for Structured {NLP} Tasks without Finetuning}. In \bibinfo{booktitle}{\emph{Proceedings of the 2023 Conference on Empirical Methods in Natural Language Processing}}, \bibfield{editor}{\bibinfo{person}{Houda Bouamor}, \bibinfo{person}{Juan Pino}, {and} \bibinfo{person}{Kalika Bali}} (Eds.). \bibinfo{publisher}{Association for Computational Linguistics}, \bibinfo{address}{Singapore}, \bibinfo{pages}{10932--10952}.
\newblock
\href{https://doi.org/10.18653/v1/2023.emnlp-main.674}{doi:\nolinkurl{10.18653/v1/2023.emnlp-main.674}}


\bibitem[Hugenroth et~al\mbox{.}(2026)]%
        {hugenroth2026critical}
\bibfield{author}{\bibinfo{person}{Philipp Hugenroth}, \bibinfo{person}{Valdemar Danry}, {and} \bibinfo{person}{Pattie Maes}.} \bibinfo{year}{2026}\natexlab{}.
\newblock \bibinfo{title}{Critical Inker: Scaffolding Critical Thinking in {AI}-Assisted Writing Through {Socratic} Questioning}.
\newblock \bibinfo{howpublished}{arXiv:2604.07167}.
\newblock
\href{https://doi.org/10.48550/arXiv.2604.07167}{doi:\nolinkurl{10.48550/arXiv.2604.07167}}


\bibitem[Jakesch et~al\mbox{.}(2023)]%
        {jakesch2023co}
\bibfield{author}{\bibinfo{person}{Maurice Jakesch}, \bibinfo{person}{Advait Bhat}, \bibinfo{person}{Daniel Buschek}, \bibinfo{person}{Lior Zalmanson}, {and} \bibinfo{person}{Mor Naaman}.} \bibinfo{year}{2023}\natexlab{}.
\newblock \showarticletitle{Co-writing with opinionated language models affects users’ views}. In \bibinfo{booktitle}{\emph{Proceedings of the 2023 CHI conference on human factors in computing systems}}. \bibinfo{pages}{1--15}.
\newblock


\bibitem[Jansen et~al\mbox{.}(2025)]%
        {jansen2025feedback}
\bibfield{author}{\bibinfo{person}{Thorben Jansen}, \bibinfo{person}{Andrea Horbach}, {and} \bibinfo{person}{Jennifer Meyer}.} \bibinfo{year}{2025}\natexlab{}.
\newblock \showarticletitle{Feedback from Generative {AI}: Correlates of Student Engagement in Text Revision from 655 Classes from Primary and Secondary School}. In \bibinfo{booktitle}{\emph{Proceedings of the 15th International Learning Analytics and Knowledge Conference}}. \bibinfo{publisher}{ACM}, \bibinfo{pages}{831--836}.
\newblock
\href{https://doi.org/10.1145/3706468.3706494}{doi:\nolinkurl{10.1145/3706468.3706494}}


\bibitem[Khot et~al\mbox{.}(2023)]%
        {khot2023decomposed}
\bibfield{author}{\bibinfo{person}{Tushar Khot}, \bibinfo{person}{Harsh Trivedi}, \bibinfo{person}{Matthew Finlayson}, \bibinfo{person}{Yao Fu}, \bibinfo{person}{Kyle Richardson}, \bibinfo{person}{Peter Clark}, {and} \bibinfo{person}{Ashish Sabharwal}.} \bibinfo{year}{2023}\natexlab{}.
\newblock \showarticletitle{Decomposed Prompting: A Modular Approach for Solving Complex Tasks}. In \bibinfo{booktitle}{\emph{The Eleventh International Conference on Learning Representations (ICLR 2023)}}.
\newblock
\urldef\tempurl%
\url{https://openreview.net/forum?id=_nGgzQjzaRy}
\showURL{%
\tempurl}


\bibitem[Kojima et~al\mbox{.}(2022)]%
        {kojima2022zeroshot}
\bibfield{author}{\bibinfo{person}{Takeshi Kojima}, \bibinfo{person}{Shixiang~Shane Gu}, \bibinfo{person}{Machel Reid}, \bibinfo{person}{Yutaka Matsuo}, {and} \bibinfo{person}{Yusuke Iwasawa}.} \bibinfo{year}{2022}\natexlab{}.
\newblock \showarticletitle{Large Language Models are Zero-Shot Reasoners}. In \bibinfo{booktitle}{\emph{Advances in Neural Information Processing Systems}}, \bibfield{editor}{\bibinfo{person}{S.~Koyejo}, \bibinfo{person}{S.~Mohamed}, \bibinfo{person}{A.~Agarwal}, \bibinfo{person}{D.~Belgrave}, \bibinfo{person}{K.~Cho}, {and} \bibinfo{person}{A.~Oh}} (Eds.), Vol.~\bibinfo{volume}{35}. \bibinfo{publisher}{Curran Associates, Inc.}, \bibinfo{pages}{22199--22213}.
\newblock
\href{https://doi.org/10.52202/068431-1613}{doi:\nolinkurl{10.52202/068431-1613}}


\bibitem[Lee et~al\mbox{.}(2025)]%
        {lee2025impact}
\bibfield{author}{\bibinfo{person}{Hao-Ping Lee}, \bibinfo{person}{Advait Sarkar}, \bibinfo{person}{Lev Tankelevitch}, \bibinfo{person}{Ian Drosos}, \bibinfo{person}{Sean Rintel}, \bibinfo{person}{Richard Banks}, {and} \bibinfo{person}{Nicholas Wilson}.} \bibinfo{year}{2025}\natexlab{}.
\newblock \showarticletitle{The Impact of Generative AI on Critical Thinking: Self-Reported Reductions in Cognitive Effort and Confidence Effects From a Survey of Knowledge Workers}. In \bibinfo{booktitle}{\emph{Proceedings of the 2025 CHI Conference on Human Factors in Computing Systems}}. \bibinfo{publisher}{Association for Computing Machinery}.
\newblock
\href{https://doi.org/10.1145/3706598.3713778}{doi:\nolinkurl{10.1145/3706598.3713778}}


\bibitem[Lee et~al\mbox{.}(2022)]%
        {lee2022coauthor}
\bibfield{author}{\bibinfo{person}{Mina Lee}, \bibinfo{person}{Percy Liang}, {and} \bibinfo{person}{Qian Yang}.} \bibinfo{year}{2022}\natexlab{}.
\newblock \showarticletitle{{CoAuthor}: Designing a Human-{AI} Collaborative Writing Dataset for Exploring Language Model Capabilities}. In \bibinfo{booktitle}{\emph{Proceedings of the 2022 CHI Conference on Human Factors in Computing Systems}}. \bibinfo{publisher}{ACM}, \bibinfo{pages}{1--19}.
\newblock
\href{https://doi.org/10.1145/3491102.3502030}{doi:\nolinkurl{10.1145/3491102.3502030}}


\bibitem[Levine et~al\mbox{.}(2025)]%
        {levine2025chatgpt}
\bibfield{author}{\bibinfo{person}{Sarah Levine}, \bibinfo{person}{Sarah~W. Beck}, \bibinfo{person}{Chris Mah}, \bibinfo{person}{Lena Phalen}, {and} \bibinfo{person}{Jaylen Pittman}.} \bibinfo{year}{2025}\natexlab{}.
\newblock \showarticletitle{How do students use {ChatGPT} as a writing support?}
\newblock \bibinfo{journal}{\emph{Journal of Adolescent \& Adult Literacy}} \bibinfo{volume}{68}, \bibinfo{number}{5} (\bibinfo{year}{2025}), \bibinfo{pages}{445--457}.
\newblock
\href{https://doi.org/10.1002/jaal.1373}{doi:\nolinkurl{10.1002/jaal.1373}}


\bibitem[Liu et~al\mbox{.}(2026)]%
        {liu2026ministral}
\bibfield{author}{\bibinfo{person}{Alexander~H. Liu}, \bibinfo{person}{Kartik Khandelwal}, \bibinfo{person}{Sandeep Subramanian}, \bibinfo{person}{Victor Jouault}, \bibinfo{person}{Abhinav Rastogi}, \bibinfo{person}{Adrien Sad{\'e}}, \bibinfo{person}{Alan Jeffares}, \bibinfo{person}{Albert Jiang}, \bibinfo{person}{Alexandre Cahill}, \bibinfo{person}{Alexandre Gavaudan}, {et~al\mbox{.}}} \bibinfo{year}{2026}\natexlab{}.
\newblock \bibinfo{title}{Ministral 3}.
\newblock
\shownote{Technical report, Mistral AI. Checkpoint used: \url{https://huggingface.co/mistralai/Ministral-3-8B-Reasoning-2512}}.
\newblock
\showeprint[arxiv]{2601.08584}~[cs.CL]
\urldef\tempurl%
\url{https://arxiv.org/abs/2601.08584}
\showURL{%
\tempurl}


\bibitem[Matcha et~al\mbox{.}(2019)]%
        {matcha2019analytics}
\bibfield{author}{\bibinfo{person}{Wannisa Matcha}, \bibinfo{person}{Dragan Ga{\v{s}}evi{\'c}}, \bibinfo{person}{Nora'Ayu Ahmad~Uzir}, \bibinfo{person}{Jelena Jovanovi{\'c}}, {and} \bibinfo{person}{Abelardo Pardo}.} \bibinfo{year}{2019}\natexlab{}.
\newblock \showarticletitle{Analytics of learning strategies: Associations with academic performance and feedback}. In \bibinfo{booktitle}{\emph{Proceedings of the 9th International Conference on Learning Analytics \& Knowledge}}. \bibinfo{publisher}{ACM}, \bibinfo{pages}{461--470}.
\newblock
\href{https://doi.org/10.1145/3303772.3303787}{doi:\nolinkurl{10.1145/3303772.3303787}}


\bibitem[Meyer et~al\mbox{.}(2026)]%
        {meyer2026comparing}
\bibfield{author}{\bibinfo{person}{Jennifer Meyer}, \bibinfo{person}{Marlene Steinbach}, \bibinfo{person}{Ronja Schiller}, \bibinfo{person}{Ute Mertens}, \bibinfo{person}{Nils-Jonathan Schaller}, \bibinfo{person}{Andrea Horbach}, \bibinfo{person}{Johanna Fleckenstein}, \bibinfo{person}{Olaf K{\"o}ller}, \bibinfo{person}{Ren{\'e}~F. Kizilcec}, {and} \bibinfo{person}{Thorben Jansen}.} \bibinfo{year}{2026}\natexlab{}.
\newblock \showarticletitle{Comparing Teacher and {AI}-Generated Feedback in the Writing Classroom: Experimental Results from Secondary School Classrooms}. In \bibinfo{booktitle}{\emph{Proceedings of the Thirteenth ACM Conference on Learning @ Scale}}. \bibinfo{publisher}{ACM}, \bibinfo{pages}{264--276}.
\newblock
\href{https://doi.org/10.1145/3774398.3811602}{doi:\nolinkurl{10.1145/3774398.3811602}}


\bibitem[Min et~al\mbox{.}(2022)]%
        {min2022rethinking}
\bibfield{author}{\bibinfo{person}{Sewon Min}, \bibinfo{person}{Xinxi Lyu}, \bibinfo{person}{Ari Holtzman}, \bibinfo{person}{Mikel Artetxe}, \bibinfo{person}{Mike Lewis}, \bibinfo{person}{Hannaneh Hajishirzi}, {and} \bibinfo{person}{Luke Zettlemoyer}.} \bibinfo{year}{2022}\natexlab{}.
\newblock \showarticletitle{Rethinking the Role of Demonstrations: What Makes In-Context Learning Work?}. In \bibinfo{booktitle}{\emph{Proceedings of the 2022 Conference on Empirical Methods in Natural Language Processing}}, \bibfield{editor}{\bibinfo{person}{Yoav Goldberg}, \bibinfo{person}{Zornitsa Kozareva}, {and} \bibinfo{person}{Yue Zhang}} (Eds.). \bibinfo{publisher}{Association for Computational Linguistics}, \bibinfo{address}{Abu Dhabi, United Arab Emirates}, \bibinfo{pages}{11048--11064}.
\newblock
\href{https://doi.org/10.18653/v1/2022.emnlp-main.759}{doi:\nolinkurl{10.18653/v1/2022.emnlp-main.759}}


\bibitem[Mouchel et~al\mbox{.}(2023)]%
        {mouchel2023revision}
\bibfield{author}{\bibinfo{person}{Luca Mouchel}, \bibinfo{person}{Thiemo Wambsganss}, \bibinfo{person}{Paola Mejia-Domenzain}, {and} \bibinfo{person}{Tanja K{\"a}ser}.} \bibinfo{year}{2023}\natexlab{}.
\newblock \showarticletitle{Understanding Revision Behavior in Adaptive Writing Support Systems for Education}. In \bibinfo{booktitle}{\emph{Proceedings of the 16th International Conference on Educational Data Mining}}. \bibinfo{publisher}{International Educational Data Mining Society}, \bibinfo{pages}{445--452}.
\newblock
\href{https://doi.org/10.5281/zenodo.8115766}{doi:\nolinkurl{10.5281/zenodo.8115766}}


\bibitem[Namaziandost and Teng(2026)]%
        {namaziandost2026socratic}
\bibfield{author}{\bibinfo{person}{Ehsan Namaziandost} {and} \bibinfo{person}{Mark~Feng Teng}.} \bibinfo{year}{2026}\natexlab{}.
\newblock \showarticletitle{{AI}-Supported {Socratic} Dialogue Agents vs. Generic {AI} Chatbots for Intermediate {EFL} Learners' Critical Thinking, Epistemic Beliefs, and Argumentation: Mixed-Methods Evidence}.
\newblock \bibinfo{journal}{\emph{International Journal of Applied Linguistics}} (\bibinfo{year}{2026}).
\newblock
\shownote{Online first}.
\newblock
\href{https://doi.org/10.1111/ijal.70268}{doi:\nolinkurl{10.1111/ijal.70268}}


\bibitem[Noy and Zhang(2023)]%
        {noy2023experimental}
\bibfield{author}{\bibinfo{person}{Shakked Noy} {and} \bibinfo{person}{Whitney Zhang}.} \bibinfo{year}{2023}\natexlab{}.
\newblock \showarticletitle{Experimental evidence on the productivity effects of generative artificial intelligence}.
\newblock \bibinfo{journal}{\emph{Science}} \bibinfo{volume}{381}, \bibinfo{number}{6654} (\bibinfo{year}{2023}), \bibinfo{pages}{187--192}.
\newblock
\href{https://doi.org/10.1126/science.adh2586}{doi:\nolinkurl{10.1126/science.adh2586}}


\bibitem[Paul and Elder(2007)]%
        {paul2007critical}
\bibfield{author}{\bibinfo{person}{Richard Paul} {and} \bibinfo{person}{Linda Elder}.} \bibinfo{year}{2007}\natexlab{}.
\newblock \showarticletitle{Critical thinking: The art of Socratic questioning}.
\newblock \bibinfo{journal}{\emph{Journal of developmental education}} \bibinfo{volume}{31}, \bibinfo{number}{1} (\bibinfo{year}{2007}), \bibinfo{pages}{36}.
\newblock


\bibitem[Pretorius and Thewissen(2026)]%
        {pretorius2026mediating}
\bibfield{author}{\bibinfo{person}{Marilize Pretorius} {and} \bibinfo{person}{Jennifer Thewissen}.} \bibinfo{year}{2026}\natexlab{}.
\newblock \showarticletitle{Mediating {L2} writing revision: The impact of teacher-moderated {ChatGPT} feedback on {L2} accuracy and lexical diversity}.
\newblock \bibinfo{journal}{\emph{Journal of Second Language Writing}}  \bibinfo{volume}{71} (\bibinfo{year}{2026}), \bibinfo{pages}{101286}.
\newblock
\href{https://doi.org/10.1016/j.jslw.2026.101286}{doi:\nolinkurl{10.1016/j.jslw.2026.101286}}


\bibitem[Scardamalia et~al\mbox{.}(1984)]%
        {scardamalia1984teachability}
\bibfield{author}{\bibinfo{person}{Marlene Scardamalia}, \bibinfo{person}{Carl Bereiter}, {and} \bibinfo{person}{Rosanne Steinbach}.} \bibinfo{year}{1984}\natexlab{}.
\newblock \showarticletitle{Teachability of reflective processes in written composition}.
\newblock \bibinfo{journal}{\emph{Cognitive science}} \bibinfo{volume}{8}, \bibinfo{number}{2} (\bibinfo{year}{1984}), \bibinfo{pages}{173--190}.
\newblock


\bibitem[Stab and Gurevych(2017)]%
        {stab-gurevych-2017-parsing}
\bibfield{author}{\bibinfo{person}{Christian Stab} {and} \bibinfo{person}{Iryna Gurevych}.} \bibinfo{year}{2017}\natexlab{}.
\newblock \showarticletitle{Parsing Argumentation Structures in Persuasive Essays}.
\newblock \bibinfo{journal}{\emph{Computational Linguistics}} \bibinfo{volume}{43}, \bibinfo{number}{3} (\bibinfo{date}{Sept.} \bibinfo{year}{2017}), \bibinfo{pages}{619--659}.
\newblock
\href{https://doi.org/10.1162/COLI_a_00295}{doi:\nolinkurl{10.1162/COLI_a_00295}}


\bibitem[Tam et~al\mbox{.}(2024)]%
        {tam2024format}
\bibfield{author}{\bibinfo{person}{Zhi~Rui Tam}, \bibinfo{person}{Cheng-Kuang Wu}, \bibinfo{person}{Yi-Lin Tsai}, \bibinfo{person}{Chieh-Yen Lin}, \bibinfo{person}{Hung-yi Lee}, {and} \bibinfo{person}{Yun-Nung Chen}.} \bibinfo{year}{2024}\natexlab{}.
\newblock \showarticletitle{Let Me Speak Freely? A Study on the Impact of Format Restrictions on Large Language Model Performance}. In \bibinfo{booktitle}{\emph{Proceedings of the 2024 Conference on Empirical Methods in Natural Language Processing: Industry Track}}, \bibfield{editor}{\bibinfo{person}{Franck Dernoncourt}, \bibinfo{person}{Daniel Preo{\c{t}}iuc-Pietro}, {and} \bibinfo{person}{Anastasia Shimorina}} (Eds.). \bibinfo{publisher}{Association for Computational Linguistics}, \bibinfo{address}{Miami, Florida, US}, \bibinfo{pages}{1218--1236}.
\newblock
\href{https://doi.org/10.18653/v1/2024.emnlp-industry.91}{doi:\nolinkurl{10.18653/v1/2024.emnlp-industry.91}}


\bibitem[Tankelevitch et~al\mbox{.}(2024)]%
        {tankelevitch2024metacognitive}
\bibfield{author}{\bibinfo{person}{Lev Tankelevitch}, \bibinfo{person}{Viktor Kewenig}, \bibinfo{person}{Auste Simkute}, \bibinfo{person}{Ava~Elizabeth Scott}, \bibinfo{person}{Advait Sarkar}, \bibinfo{person}{Abigail Sellen}, {and} \bibinfo{person}{Sean Rintel}.} \bibinfo{year}{2024}\natexlab{}.
\newblock \showarticletitle{The metacognitive demands and opportunities of generative AI}. In \bibinfo{booktitle}{\emph{Proceedings of the 2024 CHI conference on human factors in computing systems}}. \bibinfo{pages}{1--24}.
\newblock


\bibitem[Toulmin(2003)]%
        {toulmin2003uses}
\bibfield{author}{\bibinfo{person}{Stephen~E Toulmin}.} \bibinfo{year}{2003}\natexlab{}.
\newblock \bibinfo{booktitle}{\emph{The uses of argument}}.
\newblock \bibinfo{publisher}{Cambridge university press}.
\newblock


\bibitem[Wachsmuth et~al\mbox{.}(2017)]%
        {wachsmuth-etal-2017-computational}
\bibfield{author}{\bibinfo{person}{Henning Wachsmuth}, \bibinfo{person}{Nona Naderi}, \bibinfo{person}{Yufang Hou}, \bibinfo{person}{Yonatan Bilu}, \bibinfo{person}{Vinodkumar Prabhakaran}, \bibinfo{person}{Tim~Alberdingk Thijm}, \bibinfo{person}{Graeme Hirst}, {and} \bibinfo{person}{Benno Stein}.} \bibinfo{year}{2017}\natexlab{}.
\newblock \showarticletitle{Computational Argumentation Quality Assessment in Natural Language}. In \bibinfo{booktitle}{\emph{Proceedings of the 15th Conference of the {E}uropean Chapter of the Association for Computational Linguistics: Volume 1, Long Papers}}, \bibfield{editor}{\bibinfo{person}{Mirella Lapata}, \bibinfo{person}{Phil Blunsom}, {and} \bibinfo{person}{Alexander Koller}} (Eds.). \bibinfo{publisher}{Association for Computational Linguistics}, \bibinfo{address}{Valencia, Spain}, \bibinfo{pages}{176--187}.
\newblock
\urldef\tempurl%
\url{https://aclanthology.org/E17-1017/}
\showURL{%
\tempurl}


\bibitem[Wambsganss et~al\mbox{.}(2021)]%
        {wambsganss2021arguetutor}
\bibfield{author}{\bibinfo{person}{Thiemo Wambsganss}, \bibinfo{person}{Tobias K{\"u}ng}, \bibinfo{person}{Matthias S{\"o}llner}, {and} \bibinfo{person}{Jan~Marco Leimeister}.} \bibinfo{year}{2021}\natexlab{}.
\newblock \showarticletitle{{ArgueTutor}: An Adaptive Dialog-Based Learning System for Argumentation Skills}. In \bibinfo{booktitle}{\emph{Proceedings of the 2021 CHI Conference on Human Factors in Computing Systems}}. \bibinfo{publisher}{ACM}, \bibinfo{pages}{1--13}.
\newblock
\href{https://doi.org/10.1145/3411764.3445781}{doi:\nolinkurl{10.1145/3411764.3445781}}


\bibitem[Wang(2026)]%
        {wang2026question}
\bibfield{author}{\bibinfo{person}{Li-Jen Wang}.} \bibinfo{year}{2026}\natexlab{}.
\newblock \showarticletitle{Question-only AI Socratic dialogue as dialogic feedback in L2 argumentative writing: A quasi-experimental study}.
\newblock \bibinfo{journal}{\emph{Journal of Second Language Writing}}  \bibinfo{volume}{73} (\bibinfo{year}{2026}), \bibinfo{pages}{101331}.
\newblock


\bibitem[Wei et~al\mbox{.}(2022)]%
        {wei2022cot}
\bibfield{author}{\bibinfo{person}{Jason Wei}, \bibinfo{person}{Xuezhi Wang}, \bibinfo{person}{Dale Schuurmans}, \bibinfo{person}{Maarten Bosma}, \bibinfo{person}{Brian Ichter}, \bibinfo{person}{Fei Xia}, \bibinfo{person}{Ed~H. Chi}, \bibinfo{person}{Quoc~V. Le}, {and} \bibinfo{person}{Denny Zhou}.} \bibinfo{year}{2022}\natexlab{}.
\newblock \showarticletitle{Chain-of-Thought Prompting Elicits Reasoning in Large Language Models}. In \bibinfo{booktitle}{\emph{Advances in Neural Information Processing Systems}}, \bibfield{editor}{\bibinfo{person}{S.~Koyejo}, \bibinfo{person}{S.~Mohamed}, \bibinfo{person}{A.~Agarwal}, \bibinfo{person}{D.~Belgrave}, \bibinfo{person}{K.~Cho}, {and} \bibinfo{person}{A.~Oh}} (Eds.), Vol.~\bibinfo{volume}{35}. \bibinfo{publisher}{Curran Associates, Inc.}, \bibinfo{pages}{24824--24837}.
\newblock
\href{https://doi.org/10.52202/068431-1800}{doi:\nolinkurl{10.52202/068431-1800}}


\bibitem[Wood et~al\mbox{.}(1976)]%
        {wood1976scaffold}
\bibfield{author}{\bibinfo{person}{David Wood}, \bibinfo{person}{Jerome~S. Bruner}, {and} \bibinfo{person}{Gail Ross}.} \bibinfo{year}{1976}\natexlab{}.
\newblock \showarticletitle{The Role of Tutoring in Problem Solving}.
\newblock \bibinfo{journal}{\emph{Journal of Child Psychology and Psychiatry}} \bibinfo{volume}{17}, \bibinfo{number}{2} (\bibinfo{year}{1976}), \bibinfo{pages}{89--100}.
\newblock
\href{https://doi.org/10.1111/j.1469-7610.1976.tb00381.x}{doi:\nolinkurl{10.1111/j.1469-7610.1976.tb00381.x}}


\bibitem[Wu et~al\mbox{.}(2022)]%
        {wu2022aichains}
\bibfield{author}{\bibinfo{person}{Tongshuang Wu}, \bibinfo{person}{Michael Terry}, {and} \bibinfo{person}{Carrie~J. Cai}.} \bibinfo{year}{2022}\natexlab{}.
\newblock \showarticletitle{{AI} Chains: Transparent and Controllable Human-{AI} Interaction by Chaining Large Language Model Prompts}. In \bibinfo{booktitle}{\emph{Proceedings of the 2022 CHI Conference on Human Factors in Computing Systems}} (New Orleans, LA, USA) \emph{(\bibinfo{series}{CHI '22})}. \bibinfo{publisher}{Association for Computing Machinery}, \bibinfo{address}{New York, NY, USA}, \bibinfo{numpages}{22}~pages.
\newblock
\showISBNx{9781450391573}
\href{https://doi.org/10.1145/3491102.3517582}{doi:\nolinkurl{10.1145/3491102.3517582}}


\bibitem[Zhang et~al\mbox{.}(2023)]%
        {zhang2023visar}
\bibfield{author}{\bibinfo{person}{Zheng Zhang}, \bibinfo{person}{Jie Gao}, \bibinfo{person}{Ranjodh~Singh Dhaliwal}, {and} \bibinfo{person}{Toby Jia-Jun Li}.} \bibinfo{year}{2023}\natexlab{}.
\newblock \showarticletitle{{VISAR}: A Human-{AI} Argumentative Writing Assistant with Visual Programming and Rapid Draft Prototyping}. In \bibinfo{booktitle}{\emph{Proceedings of the 36th Annual ACM Symposium on User Interface Software and Technology}}. \bibinfo{publisher}{ACM}, \bibinfo{pages}{1--30}.
\newblock
\href{https://doi.org/10.1145/3586183.3606800}{doi:\nolinkurl{10.1145/3586183.3606800}}


\bibitem[Zhu et~al\mbox{.}(2019)]%
        {zhu2019keystroke}
\bibfield{author}{\bibinfo{person}{Mengxiao Zhu}, \bibinfo{person}{Mo Zhang}, {and} \bibinfo{person}{Paul Deane}.} \bibinfo{year}{2019}\natexlab{}.
\newblock \showarticletitle{Analysis of keystroke sequences in writing logs}.
\newblock \bibinfo{journal}{\emph{ETS Research Report Series}} \bibinfo{volume}{2019}, \bibinfo{number}{1} (\bibinfo{year}{2019}), \bibinfo{pages}{1--16}.
\newblock
\href{https://doi.org/10.1002/ets2.12247}{doi:\nolinkurl{10.1002/ets2.12247}}


\end{thebibliography}

\end{document}